\documentclass[aps,prx,twocolumn,superscriptaddress,longbibliography,eqsecnum]{revtex4-2}
\usepackage{amsmath,amssymb}
\usepackage[pdftex]{hyperref,graphicx}
\hypersetup{colorlinks = true, urlcolor = blue, linkcolor = blue, citecolor = blue}
\usepackage{physics}
\usepackage{xcolor,ulem}
\usepackage{bm}
\usepackage{dsfont}

\newcommand{\mC}{{\bm{m}_{\mathcal{C}}}}
\newcommand{\C}{{\mathcal{C}}}
\newcommand{\HPadd}[1]{{\leavevmode\color{black}{#1}}}

\newcommand{\jp}[1]{\textcolor{black}{#1}}

\begin{document}
\title{Control-driven critical fluctuations across quantum trajectories}
\author{Haining Pan}
\affiliation{Department of Physics and Astronomy, Center for Materials Theory, Rutgers University, Piscataway, NJ 08854 USA}
\author{Thomas Iadecola}
\affiliation{Department of Physics and Astronomy, Iowa State University, Ames, IA 50011, USA}
\affiliation{Ames National Laboratory, Ames, IA 50011, USA}
\affiliation{Department of Physics, The Pennsylvania State University, University Park, Pennsylvania 16802, USA}
\affiliation{Institute for Computational and Data Sciences, The Pennsylvania State University, University Park, Pennsylvania 16802, USA}
\affiliation{Materials Research Institute, The Pennsylvania State University, University Park, Pennsylvania 16802, USA}
\author{E. M. Stoudenmire}
\affiliation{Center for Computational Quantum Physics, Flatiron Institute, New York, New York 10010, USA}

\author{J. H. Pixley}
\affiliation{Department of Physics and Astronomy, Center for Materials Theory, Rutgers University, Piscataway, NJ 08854 USA}
\affiliation{Center for Computational Quantum Physics, Flatiron Institute, New York, New York 10010, USA}
\date{\today}

\begin{abstract}
Monitored quantum circuits in which entangling unitary dynamics compete with projective local measurements can host measurement-induced phase transitions witnessed by entanglement measures at late times.
Adding feedback conditioned on the measurement outcomes gives rise to another type of phase transition witnessed by local order parameters and correlation functions. 
These transitions, known as control or absorbing-state transitions, generically occur within the area-law entanglement phase and are thought to be governed by classical physics in that their critical exponents match those of the classical limit of the model. 
In this work, we examine quantum features of these transitions, focusing on a Bernoulli circuit model with a well-defined classical limit, in the presence of unitary dynamics, measurements, and feedback.
First we demonstrate that, in the local basis defined by the absorbing state, the steady-state quantum coherence undergoes a phase transition at the control transition, where its logarithm changes discontinuously from volume- to area-law scaling.
Second, we analyze the control transition from the perspective of fluctuations in observables, which carry two contributions: classical fluctuations over circuit realizations (present in the classical limit), and quantum fluctuations over trajectories and states (both absent in the classical limit).
Both contributions can be estimated in experiments without post-selection.
The circuit-to-circuit fluctuations, the dominant contribution, carry the critical behavior of the classical limit. 
However, the subleading quantum fluctuations that represent fluctuations between different quantum ``worlds'' also go critical at the control transition.
These critical quantum fluctuations at the control transition also occur in other models, and we discuss how they can be measured experimentally without post-selection. 
\end{abstract}
\maketitle

\section{Introduction}\label{sec:introduction}
  
The non-equilibrium dynamics of strongly interacting quantum systems is a fundamental and timely topic at the intersection of quantum information science and many-body physics.
In addition to entangling unitary dynamics, tasks like quantum error correction and state preparation require the ability to perform mid-circuit measurements (i.e., to measure local parts of the system in real time).
This capability, which is becoming available on a variety of noisy intermediate-scale quantum (NISQ) hardware platforms, will be an important driver of the transition to the early fault-tolerant era of quantum computing.

The competition between projective local measurements and entangling unitary dynamics drives a measurement-induced phase transition (MIPT) between dynamical phases with extensive volume-law and subextensive area-law entanglement at late times~\cite{skinner2019measurementinduced,li2018quantum,li2019measurementdriven, chan2019unitaryprojective,fisher2023random}.
This phenomenon is inherently quantum mechanical as it can only be witnessed in quantities (e.g. any entanglement diagnostic) that are non-linear in the reduced density matrix.
This is a consequence of the fact that the reduced density matrix becomes featureless upon averaging over the outcomes of the projective measurements at any measurement rate.
To circumvent this, entanglement measures must be computed in a manner that resolves the quantum trajectories associated with different measurement histories.
While this can be achieved experimentally for small systems~\cite{noel2022measurementinduced,hoke2023measurementinduced}, the number of measurement histories scales exponentially in the number of measurements, which makes scaling up prohibitive.
Despite this challenge, known as the post-selection problem, experiments have now observed signatures of the MIPT on trapped ion and superconducting qubit NISQ devices.
This defines a new notion of dynamical phases of quantum matter and non-equilibrium phase transitions with universality classes that share close connections to percolation physics.

A natural generalization of measurement-enriched quantum dynamics is adaptive quantum dynamics, where the measurement outcomes are used to {control the subsequent dynamics.  In particular, we will consider models where they are used to} steer the dynamics towards a pre-determined state.  For this target state to be stable, generic unitary dynamics cannot be used; instead, a dark or control state is embedded into the unitary dynamics that can allow the feedback from the measurement outcomes to overcome the quantum chaotic evolution.  As a result, a control-induced phase transition (CIPT) has been uncovered in several contexts~\cite{iadecola2023measurement,lemaire2024separate,pan2024local,piroli2023triviality, sierant2023controlling,odea2024entanglement,ravindranath2023entanglement,thompson2024population}, which has now been observed on NISQ devices of trapped ions~\cite{chertkov2023characterizing} and superconducting qubits~\cite{pokharel2025order}.  
This is a transition between an active phase, where the system's steady state is by some measures far from the control state; and a control phase, where the system's steady state is either a dark state with no fluctuations or is near the control state with very limited fluctuations.
This defines two classes of models possessing CIPTs. 
The first can be directly constructed from the unstable orbits of a classically chaotic dynamical system; as a result, such a state is not an exact dark state in the quantum limit.
The second class has an exact dark state that is built into the dynamics without any reference to a classical system.
In this work, we will first use the Bernoulli circuit model~\cite{iadecola2023measurement,lemaire2024separate,allocca2024statistical,pan2024local} under the first class to demonstrate the quantum nature of the CIPT, and demonstrate that the same quantum features also extend to the second class of models using two examples with an exact dark state~\cite{odea2024entanglement, sierant2023controlling}.
We note that the CIPT is also often referred to as an absorbing-state transition, a term borrowed from classical nonequilibrium statistical mechanics~\cite{henkel2008nonequilibrium}, and in this work we will use these two terms interchangeably.

Importantly, unlike feedback-free MIPTs, CIPTs are naturally witnessed by local order parameters or correlation functions that are linear in the density matrix, {so they can be studied without postselection.}  However, a CIPT may or may not coincide with a transition out of the volume-law phase (i.e., a MIPT) depending on whether the feedback acts locally or globally on the quantum state; generically, the MIPT occurs within the active phase, preceding the CIPT, but global feedback can drive them together~\cite{pan2024local,piroli2023triviality,odea2024entanglement,sierant2023controlling}.
At the classical level, these systems realize dynamical phase transitions that result from either the stochastic or deterministic control of chaotic evolution.
Embedding such classical critical dynamics into a quantum system is a systematic protocol that can induce a CIPT in a monitored quantum many-body system with feedback.  In this paper, we only consider dynamics that take pure states to pure states, so there is no unmonitored coupling of the system to its environment that would produce a mixed state of the system.

As a result, a central question in the field, which is also a main focus of the present manuscript, is if this dynamical transition remains purely classical or if there are any inherently universal, quantum critical degrees of freedom present due to the interplay of measurements, feedback, and chaotic unitary dynamics.

To make this question precise, it is helpful to first identify what kinds of quantum fluctuations are possible in adaptive quantum circuits.
The first, which is shared by unitary quantum many-body systems, are the quantum uncertainties that are present whenever there is quantum entanglement in the state of the system.
The second, which is due to measurements and possible feedback, involves the differences between quantum states that are produced by different quantum trajectories (different measurement outcomes).
In other words, the latter involves fluctuations across the many possible ``worlds'' of quantum mechanics, corresponding to different measurement and feedback histories.  In monitored quantum circuits without feedback the trajectory-averaged density matrix, which includes both types of fluctuations, goes to a long-time steady state that maximizes the entropy, and thus the MIPT is not detectable in quantities that are linear in the density matrix.
However, introducing feedback may drive the system to critical behavior (such as the CIPT) that is detectable in {the trajectory-averaged density matrix, and thus without postselection.}

In this work, we provide precise definitions of classical and quantum fluctuations, including both coherent (within trajectory) and incoherent (across quantum trajectories or ``worlds'') quantum contributions, which we use to investigate the quantum properties of CIPTs.  The critical universality classes uncovered in quantum CIPTs have so far been governed by classical stochastic processes:
a random walk in Ref.~\cite{iadecola2023measurement}, directed percolation in Refs.~\cite{piroli2023triviality, sierant2023controlling,odea2024entanglement}, a classical branching-annihilating random walk in Ref.~\cite{ravindranath2023entanglement}, and a probabilistic cellular automaton with long-range correlations in Ref.~\cite{iadecola2024concomitant}.
We stress that while in each of these cases, the entanglement does display a transition (either volume-to-area or area-to-zero) at the CIPT, it does so with the critical exponents of the classical model. 
As a result, the universal properties of the entanglement do not simply unveil the distinction between classical and quantum contributions to the critical properties of the CIPT and new insight is required.
In the following, we provide a framework to quantitatively describe the universal quantum fluctuations at CIPTs, transitions that arise from embedding a classical dynamical transition into a quantum system, which is applicable to a broad class of adaptive random quantum circuits undergoing unitary dynamics, measurements, and feedback.

To unveil the quantum features of the CIPT we focus on a model that embeds the classical Bernoulli map under stochastic control~\cite{antoniou1996probabilistic,antoniou1997probabilistic,antoniou1998absolute} into a qubit chain with Haar random unitary gates and conditional reset operations. 
To show that the framework we present within \HPadd{applies to other adaptive quantum circuits, we also analyze two quantum models with CIPTs that have an exact dark state in the dynamics in Sec.~\ref{sec:generalization}: a Haar random adaptive circuit in Ref.~\onlinecite{odea2024entanglement} and a bricklayer Clifford random adaptive monitored circuit in Ref.~\onlinecite{sierant2023controlling}.}
We also point out that these quantum effects can be described by mapping to a classical statistical mechanics model in one higher dimension~\cite{allocca2024statistical}, though distinct quantum observables,  entanglement metrics, and distinct trajectory averages can each require their own classical models.

This ``Bernoulli circuit" model~\cite{iadecola2023measurement,lemaire2024separate,allocca2024statistical,pan2024local} is ideal for our purposes for several reasons. 
First, it has been studied in both classical and quantum circuits in Refs.~\cite{iadecola2023measurement, pan2024local} and the phase diagram is well understood. 
Second, the classical model has a control transition that can be derived analytically and whose universal properties are described by a simple random walk. 
Third, the model's dynamics are naturally described in terms of an emergent quasiparticle, the first domain wall (defined in Sec.~\ref{sec:overview}), that is a sharp point-like object in the classical model but becomes a wave packet in the quantum model; it is this object that undergoes an unbiased random walk at the CIPT.
Last, by focusing on local feedback, we can split the CIPT and MIPT so that the control transition takes place entirely within the area-law phase.
This enables an efficient simulation using matrix product states (MPS), allowing us to study the critical behavior of the CIPT in systems of up to $40-60$ sites with periodic boundary conditions (see Appendix~\ref{app:FSS} for a better estimate of critical exponents).

To set the stage, we begin by considering the probability density of the quantum state averaged over samples (including random initial states, circuit realizations, and measurement outcomes), finding that this ``average wave function" inherits the behavior of the classical model.
We, therefore, turn to study the fluctuations of the wave function, which are comprised of both classical and quantum fluctuations.
If we do not distinguish between these contributions, the classical behavior dominates the signal. 
However, if we carefully separate out the quantum fluctuations arising from random measurement outcomes and quantum state superpositions and analyze them in their own right, we find that they become critical with universal scaling properties captured by a scaling dimension that is zero classically but nonzero in the quantum limit. 
To show these quantum fluctuations are phase coherent we compute the ensemble-averaged $l_1$-coherence~\cite{baumgratz2014quantifying} of the quantum state (defined below) \HPadd{for a Bernoulli circuit model}, which is quite challenging to compute using MPS. 
Nonetheless, we are able to compute the $l_1$-coherence accurately by utilizing the tensor cross interpolation approach~\cite{oseledets2010ttcross,dolgov2020parallel,ritter2024quantics,fernandez2024learning}, which allows us to learn the mapping from an individual bit string to the norm of the corresponding wave function amplitude in the computational basis.
The complexity of the method is $O(\chi^4)$, where $\chi$ is the maximal bond dimension of the MPS. 
This allows us to definitively show that the CIPT in the Bernoulli circuit model is, in fact, a transition from quantum coherent to classical dynamics. 
\HPadd{Beyond the Bernoulli circuit model, the exponential growth of the quantum coherence also manifests in the Haar random adaptive circuit, and bricklayer Clifford random adaptive monitored circuit, where the latter runs up to 256 qubits.}

The rest of the paper is organized as follows.
In Sec.~\ref{sec:overview}, we present the model, define the classical and quantum fluctuations, and outline the main results of this work. 
In Sec.~\ref{sec:mean}, we present numerical evidence that the quantum CIPT inherits the features of the classical one at the level of the average wave function. 
We also probe the $l_1$-coherence across the CIPT, unveiling the transition's quantum nature.
In Sec.~\ref{sec:fluctuation}, we show that the quantum nature of the control transition manifests in the fluctuations of the wave function. 
In particular, while the circuit-to-circuit fluctuations for the classical and quantum models are both governed by the random walk universality class, the trajectory-to-trajectory fluctuations develop a non-zero scaling dimension that is universal and only non-zero in the quantum model.
\HPadd{In Sec.~\ref{sec:generalization}, we demonstrate the generality of this methodology by studying the exponential scaling of the quantum coherence, and the critical quantum fluctuations, in two adaptive quantum circuit models with an exact dark state.}
We conclude and offer an outlook in Sec.~\ref{sec:discussion}.
In Appendix~\ref{app:distribution}, we provide additional details on statistical properties of the wave function.
In Appendix~\ref{app:coherence_Haar}, we derive an expression for the coherence of a Haar random state.
In Appendix~\ref{app:TCI}, we provide technical details on estimating the coherence in the MPS simulations using tensor cross interpolation.
In Appendix~\ref{app:fluctuation_O}, we study the three kinds of fluctuations examined in this paper for another metric, the magnetization density, to demonstrate the scaling dimension we have estimated is universal.
In Appendix~\ref{app:FSS}, we use our MPS numerics to obtain a better estimate of the critical point and correlation length exponent of the CIPT studied for smaller system sizes in Ref.~\cite{pan2024local}.
\HPadd{In Appendix~\ref{app:C_Clifford}, we derive the expression for efficiently estimating the quantum coherence in the Clifford circuit model using the tableau representation.}

\section{Background and Overview of Results}\label{sec:overview}
In this section, we first introduce the Bernoulli circuit model used in this paper to demonstrate the concept of isolating the classical and quantum fluctuations in Sec.~\ref{sec:model}, and then define each type of fluctuations in detail in Sec.~\ref{sec:separating}. Finally, we outline the main results of this work in Sec.~\ref{sec:main_results}.
\subsection{Bernoulli circuit model}\label{sec:model}
\begin{figure*}[htbp]
    \centering
    \includegraphics[width=6.8in]{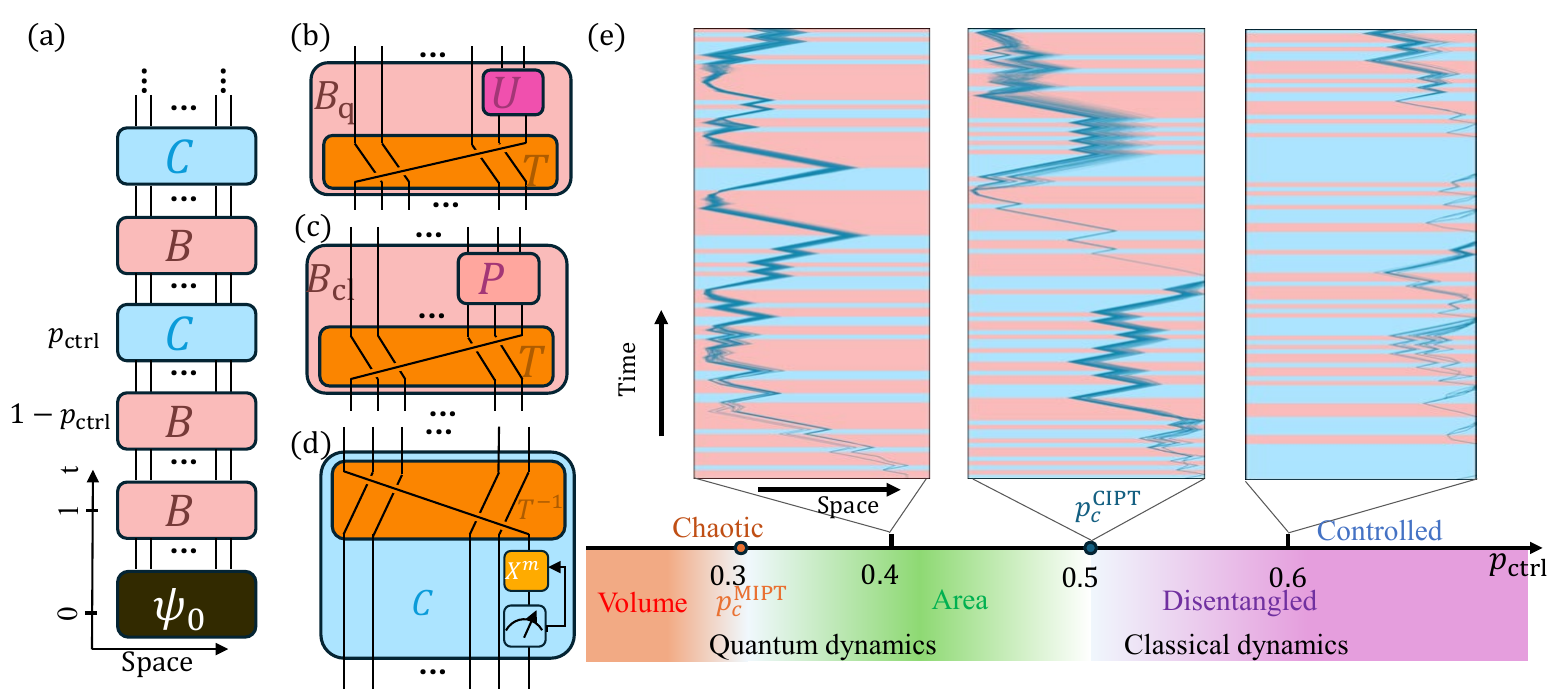}
    \caption{
    (a) Schematic of the stochastic dynamics in which, at each time step, the control map $C$ is applied with probability $p_{\text{ctrl}}$ and the Bernoulli map ($B$) is applied with probability $1-p_{\text{ctrl}}$.
    (b,c) Circuit diagrams for the quantum and classical versions of the Bernoulli map, respectively.
    (d) Circuit diagram of the control map, which is the same for both the classical and quantum models.
    (e) Controlled phase diagram indicating the CIPT at $p_{c}^{\text{CIPT}}=0.5$ separating the chaotic and controlled phases. 
    The plots show typical trajectories of the first domain wall (see Sec.~\ref{sec:FDW} for a definition) given the same quantum circuit in the chaotic (left), critical (middle), and controlled (right) phases, respectively [see Eq.~\eqref{eq:traj_fluct}].  The cyan and red strips
    indicate the application of control and Bernoulli maps, respectively.
    The bottom bar indicates the MIPT from the volume-law to the area-law phase at $p_{c}^{\text{MIPT}}\approx 0.3$. Above the CIPT, the system becomes trapped {near the control state
    and becomes disentangled in the bulk.}
    }
    \label{fig:schematic}
\end{figure*}

We begin by defining the Bernoulli circuit model as shown in Fig.~\ref{fig:schematic} and reviewing its phase diagram~\cite{iadecola2023measurement,pan2024local}.
We first describe the classical version of the model before discussing what changes when quantum superpositions are introduced.

\subsubsection{Classical Model}
The chaotic dynamics in the Bernoulli circuit model are derived from the classical Bernoulli map defined as ~\cite{renyi1957representations}
\begin{equation}\label{eq:B}
    B: x \mapsto 2x~\text{mod}~1,
\end{equation}
where $x$ is a real number between 0 and 1.
For any rational number $x_0$, the Bernoulli map generates a finite-length periodic orbit. 
However, the orbit of any rational $x_0$ is unstable under infinitesimally small deviations $x_0\to x_0+\delta x$ as the irrationals are dense on the number line, leading to chaotic dynamics.
Thus, we introduce a control map that aims to push the dynamics towards the rational numbers $\{x_f \}$ on the orbit of $x_0$.
For any $x\in [0,1)$, the control map acts as
\begin{equation}\label{eq:C}
    C:x \mapsto x_f/2 + x/2,
\end{equation}
where $x_f$ is the point on the orbit closest to $x$.
In the following, we restrict our attention to the orbit of $x_0=0$, which corresponds to a single fixed point $x_f=0$ such that $C:x\mapsto x/2$.
At each time step, the control map is stochastically applied to the system with probability $p_{\text{ctrl}}$ to counteract the chaotic dynamics generated by the Bernoulli map, which is applied with probability $1-p_{\text{ctrl}}$, as shown in Fig.~\ref{fig:schematic}(a)~\cite{antoniou1996probabilistic}.

In order to move towards a quantum analog of this model, we encode the real number $x$ in a binary representation truncated to $L$ bits,  
\begin{equation}
    \left( x \right)_{10}=\left( {0.b_1b_2\dots b_L} \right)_2=\ket{b_1b_2\dots b_L} \equiv\ket{x},
\end{equation}
where $b_i=\left\{ 0,1 \right\}$.
Here, we use the bra-ket notation, which applies to both classical and quantum models, with the only difference being that the classical model is restricted to a single element of the computational basis at any given time $\ket{b_1b_2\dots b_L}$, while the quantum model allows for superpositions of these states.
Under this mapping, the fixed point $x_f=0$ becomes the polarized state $\ket{00\dots0}$, {which is the ``control state''}.

With the discretization into bit strings, the multiplication by 2 in the Bernoulli map $B$ can be realized by a leftward shift of the entire bit string, i.e., 
\begin{equation}
    T\ket{b_1b_2\dots b_L}= \ket{b_2\dots b_Lb_1},
\end{equation}
where the first bit $b_1$ is cycled to the leftmost bit to ensure unitarity (or equivalently, we impose the periodic boundary condition).
However, this discretization does not yet accommodate chaotic dynamics since the $L$-bit binary representation at finite $L$ can only represent a rational number $\sum_{i=1}^L b_i2^{-i}\in \mathbb{Q}$ and therefore must return to its initial value after at most $L$ applications of the left-shift $T$.
Therefore, to restore the chaotic dynamics, we introduce a scrambler acting on the last three bits. 
We choose the scrambler as a random element $P\in S_8$, the permutation group of the computational basis spanned by the last three bits. 
Thus, the discretized Bernoulli map or Bernoulli circuit is realized by the composition of the left-shift and the scrambler, i.e., 
\begin{equation}
\label{eq:Bcl}
    B_{\text{cl}}=PT,
\end{equation}
as shown in Fig.~\ref{fig:schematic}(b).
We note that this formulation of the Bernoulli map captures the action of the original map on the normal numbers~\cite{emileborel1909probabilites}, which are a dense subset of the irrational numbers whose binary expansion is uniformly random.
This is sufficient to reproduce the fundamental aspects of the chaotic dynamics in the limit of large $L$, including the model's Kolmogorov-Sinai entropy of $\log 2$~\cite{renyi1957representations,lemaire2024separate}.

The control map $C$ is naturally realizable in the discretized setting.
When the control targets the fixed point $x_f=0$, $C$ reduces to division by $2$ [see Eq.~\eqref{eq:C}].
This can be accomplished by resetting the last bit to $0$, followed by a rightward cyclical shift of the entire bit string as shown in Fig.~\ref{fig:schematic}(d), i.e.,
\begin{equation}\label{eq:control}
    \begin{split}
        C \ket{b_1b_2\dots b_L}&=T^{-1} R \ket{b_1b_2\dots b_L}\\
        &=T^{-1} (X_L)^{b_L} \ket{b_1b_2\dots b_L}\\
        &=\ket{0b_1b_2\dots b_{L-1}},
    \end{split}
\end{equation}
where $X_{L}$ is the Pauli $X$ that flips the value of the bit at the last site.
Note that the last bit is flipped only if it was initially set to $1$, guaranteeing that the first bit in the string will be zero after the rightward cyclic bit shift.
This ensures that the new bit string will correspond to a number less than $1/2$.
This process erases the initial value of $b_L$, so the control map in this formulation is intrinsically nonunitary (i.e., irreversible).

\subsubsection{Quantum model}
The quantum model differs from the classical model in two important ways.
First, the scrambler in the Bernoulli map changes from a 3-bit permutation $P$ to a 2-qubit Haar random unitary $U$ acting on the last two qubits, as shown in Fig.~\ref{fig:schematic}(c). That is, the Bernoulli map in the quantum model is given by
\begin{equation}
\label{eq:Bq}
    B_{\text{q}}=UT,
\end{equation}
which, unlike Eq.~\eqref{eq:Bcl}, generates superpositions when acting on a computational basis state.
Second, the reset operation $R$ entering the control map $C$ [Eq.~\eqref{eq:control}] is implemented using a quantum measurement.
A quantum measurement of the $L$th qubit changes a generic quantum state $\ket{\psi}$ according to the nonlinear operation
\begin{equation}
    M_L(m) \ket{\psi} = \frac{P_L(m)\ket{\psi}}{\sqrt{\expval{P_L(m)}{\psi}}},
\end{equation}
where $m=0,1$ is the outcome of the measurement and 
\begin{equation}\label{eq:P}
    P_i(m) = \ketbra{m}
\end{equation}
is the projection operator onto outcome $m$ at site $i$.
The action of the reset operation $R$ in Eq.~\eqref{eq:control} on a single computational basis state can then be expressed as
\begin{equation}\label{eq:reset}
    \begin{split}
        R \ket{b_1b_2\dots b_L}
        &=(X_L)^{m} M_L(m) \ket{b_1b_2\dots b_L}\\
        &=\ket{b_1b_2\dots 0}.
    \end{split}
\end{equation}
Note that, for a single computational basis state, $m=b_L$ is the only possible measurement outcome and the above is equivalent to the reset performed in Eq.~\eqref{eq:control}.
However, for a superposition of quantum states the measurement outcomes are no longer deterministic; rather, they are distributed according to the Born rule, i.e., $m$ is drawn at random with probability ${\expval{P_L(m)}{\psi}}$ each time a measurement is performed.

\subsubsection{First domain wall}\label{sec:FDW}
A major benefit of working with the present model is that its dynamics can be captured by an object called the first domain wall (FDW).
The FDW in an individual computational basis state is defined with respect to the target state $\ket{00\dots0}$ as the location of the first ``1" in the bit string; e.g., for the state $\ket{00000\underline{1}100101000\dots}$ the location of the FDW is labeled by the underline.
The FDW can be viewed as the boundary between controlled and uncontrolled regions in the bit string. 
Its position, measured from the right edge of the bit string, quantifies the distance between that bit string and the fixed point $x_f=0$.
In particular, when the FDW is at position $k$, $x-x_f<2^{-L+k}$; the extreme limits $k=L$ and $k=0$ correspond to fully uncontrolled and fully controlled, respectively. 
The FDW also acts like an effective ``quasiparticle'' that we can track as the system evolves. 
In particular, we can consider the evolution of an initial (product) state with the FDW at position $k$: $\ket{\psi_0} = \ket{0}^{\otimes L-k}\ket{1}\bigotimes^{k-1}_{i=1}\ket{b_i}$ with arbitrary $b_i=0,1$.
In the classical model, the control map translates the FDW to the right, while the chaotic map translates it to the left, leaving behind scrambled bits in its wake.
In the quantum model, the FDW's location can be promoted to a quantum operator $\hat k$.
Under dynamics starting from a computational basis state with the FDW at $k$, the quantum mechanical average FDW position $\expval{\hat k(t)}$ constitutes an ``entanglement front" separating a disentangled region to its left from an entangled region to its right (where the Haar scrambler acts), see e.g.~Fig.~1 in Ref.~\cite{iadecola2023measurement}.
Moreover, the FDW is no longer a perfectly sharp quasiparticle; rather, it is a wave packet with mean position $\expval{\hat k(t)}$ and spread $\sigma^2_s=\expval{\hat k^2(t)}-\expval{\hat k(t)}^2$.
Thus, the FDW is a convenient object that we can use to separate out classical and quantum aspects of the dynamics.

\subsubsection{Phase diagram}
The phase diagram of the quantum Bernoulli circuit model involves both a CIPT, where the system transitions from an uncontrolled to a controlled phase, and an MIPT, where the system transitions from a volume-law to an area-law phase~\cite{iadecola2023measurement,lemaire2024separate,allocca2024statistical,pan2024local}.
The controlled and uncontrolled phases can be characterized by a local order parameter, the magnetization density
\begin{equation}\label{eq:magnetization}
    M_z=\frac{1}{L}\sum^L_{i=1}Z_i,
\end{equation}
where $Z_i$ is a Pauli operator whose eigenstates are the local computational basis states.
In the {``chaotic'' or} uncontrolled phase, the system {stays away from the control state}
and $\expval{M_z}=0$ at late times {and $L\rightarrow\infty$,} while in the controlled phase, the system always {is driven near the control state,}
and the magnetization density is nonzero at late times.  The volume- and area-law phases are characterized by the system-size scaling of entanglement measures as discussed in Sec.~\ref{sec:introduction}.

In principle, the CIPT and MIPT can occur simultaneously or separately as a function of the control rate $p_{\rm ctrl}$, depending on the locality of the control map~\cite{pan2024local}.
When the control map targets the fixed point $x_f=0$ (i.e., the polarized state $\ket{00\dots0}$), they occur separately, and three phases emerge: a volume-law chaotic phase, an area-law chaotic phase, and a disentangled controlled phase [see Fig.~\ref{fig:schematic}(e)].
The MIPT between volume-law and area-law chaotic phases occurs at $p^{\rm MIPT}_c\approx 0.3$ and its nature is inherently quantum.
The CIPT between area-law chaotic and controlled phases occurs at $p^{\rm CIPT}_{c}=0.5$ in both the classical and quantum models defined by Eqs.~\eqref{eq:Bcl} and \eqref{eq:Bq}, respectively.
For more details on the phase diagram, we refer to Refs.~\cite{iadecola2023measurement,pan2024local}.

In this paper, we are primarily interested in the features of the CIPT around $p_{\text{ctrl}}=0.5$, which occurs within the area-law phase
of the model.
Thus, in addition to the usual exact state vector evolution~\cite{iadecola2023measurement,pan2024local}, we can take advantage of matrix product states (MPS) to efficiently study the critical behavior (note that the entanglement remains area-law even at the CIPT).
We adopt a technique called tensor cross interpolation to compute the quantum coherence in polynomial complexity, which would otherwise be exponentially hard to compute in the MPS formalism.

\subsection{Separating classical and quantum fluctuations}\label{sec:separating}
\begin{figure*}[htbp]
    \centering
    \includegraphics[width=6.8in]{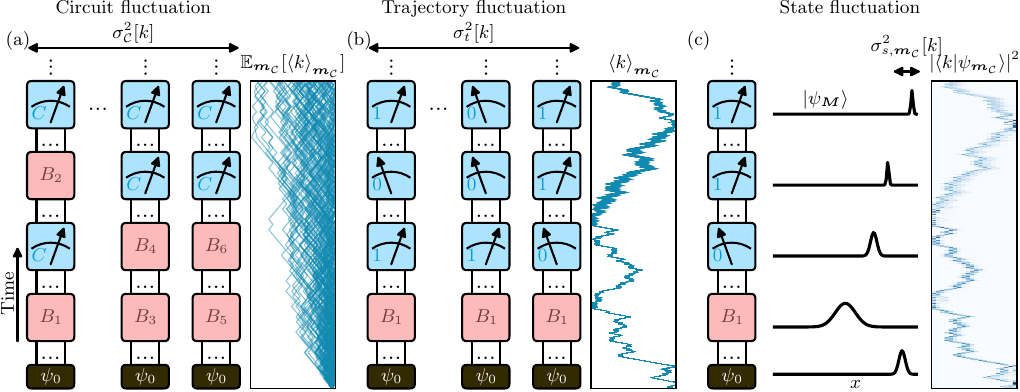}
    \caption{
    {At the critical point $p_{\text{ctrl}}=0.5$ in the Bernoulli circuit model in Sec.~\ref{sec:model}:} 
    (a) Left: Schematic of the circuit fluctuation of the first domain wall $k$ as an example {(the position of the first non-fixed point component ``1" in the wave function computational basis, see Eq.~\eqref{eq:k})}, $\sigma_{\C}^2[k]$, across different realization of Bernoulli map $B_i$ and position of control maps $C$ (see definitions in Sec.~\ref{sec:model} and Figs.~\ref{fig:schematic}(b,d)). Right: Numerical data show many realizations of the FDW dynamics for $t<4L$ {with always the same initial state with $k$ having a low uncertainty}, which sample averaging reveals to proceed diffusively~\cite{iadecola2023measurement}. {Note that in actual simulation, e.g., Fig.~\ref{fig:cl_fluct_FSS}, we choose the initial product state with $k=L/2$ such that it simulates the unbiased random walk best in $\mathbb{Z}$.}
    (b) Left: Given a fixed circuit $\C$ {from (a)}, the trajectory fluctuation $\sigma_{t}^2[k]$ across different measurement outcomes $\mC$.
    Right: Numerical data run up to $t<L^2$ indicating fluctuations around the trajectory dominated by a specific circuit.
    (c) Left: Given a fixed circuit $\C$ {from (a)} and measurement history $\mC$ {from (b)}, the state fluctuations $\sigma_{s,\mC}^2[k]$ for a single pure state $\ket{\psi_{\mC}}$, indicating the spreading of the FDW wave packet in that state.
    Right: Numerical data show the distribution of the FDW for a single circuit and measurement history with $t<L^2$.
    }
    \label{fig:schematic_fluct}
\end{figure*}

We now introduce the main concept of isolating different types of fluctuations to reveal the quantum nature of seemingly-classical transitions in the adaptive monitored circuit. 
This idea is broadly applicable to various models with a dark or control state, e.g., the CIPT in the competition between chaotic and controlled dynamics~\cite{iadecola2023measurement}, and absorbing-state transitions in adaptive monitored circuits~\cite{odea2024entanglement,sierant2023controlling}.
\HPadd{In the following results, we will first use the Bernoulli CIPT as a paradigmatic example to illustrate the concept, and then show its application to other adaptive monitored models with absorbing-state transition in Sec.~\ref{sec:generalization}.}

The aim is to determine the variance of a generic quantum operator $O$. 
To begin, we define all quantities being averaged.
Starting from an initial pure state $\ket{\psi_0}$, we have a sequence of stochastic events that either apply a unitary or a measurement with feedback.
The combination of the initial state and the sequence of unitaries and measurements defines one circuit realization labeled by $\C$. 
Each time step $t_i$ (e.g., $t=0,1,2\dots$ in Fig.~\ref{fig:schematic}(a)) that the circuit performs a measurement, the outcome is randomly sampled from the Born probability distribution with outcome $m^{(i)}$. 
We denote the full measurement history of a trajectory $\bm{m}$ for circuit realization $\C$ by the vector $\mC=\left( m^{(1)},m^{(2)},\dots \right)$; we seek to parse the fluctuations into those due to variations between circuits and those due to variations
within a fixed circuit $\C$. 

We are now in a position to specify an average over ``samples'', where each ``sample'' is a single trajectory of a single circuit, indexed by $\bm{M}=\{ \C, {\bm{m}}_{\C} \}$.
The average over samples is then 
    \begin{equation}\label{eq:decompose}
        \begin{split}
            \mathbb{E}_{\bm{M}}[\dots] =& \sum_{\bm{M}} p_{\bm{M}}[\dots] =
            \sum_{\C} p_{\C}\sum_{\mC}p_{\mC|\C}[\dots] \\
            = &\mathbb{E}_{\C} \left[ \mathbb{E}_{\mC}[\dots] \right]~.
            \end{split}
        \end{equation}
Note that since this is a simple average (no post-selection), it can be experimentally estimated by gathering many samples.
We further stress the nested form of the averages (also denoted through the conditional probability $p_{\mC|\C}$) that allows us to take the classical limit, which has deterministic measurement outcomes, so for fixed circuit $\C$,
{all classical trajectories are identical.}

{If we make projective measurements of operator $O$, giving results $O_{\bm{M}}$ for $\bm{M}$,}
the variance of these single-shot experimental observations $O_{\bm{M}}$ across $\bm{M}$ can be estimated without post-selection by {
\begin{equation}\label{eq:observable_fluctuation}
\sigma_{\bm{M}}^2[O]=\mathbb{E}_{\bm{M}} \left[O^2_{\bm{M}} \right] - \left( \mathbb{E}_{{\bm{M}}}\left[O_{\bm{M}} \right] \right)^2.
\end{equation}}
It is now straightforward, using Eq.~\eqref{eq:decompose}, to see that this will involve a combination of fluctuations over quantum trajectories (i.e., measurement outcomes for a fixed circuit realization) and circuit-to-circuit fluctuations that also exist in the classical limit of the model.
Therefore, we can now decompose this variance over samples into classical circuit fluctuations and quantum fluctuations: 

\begin{widetext}
\begin{equation}\label{eq:fluct}
\begin{split}
    \sigma_{\bm{M}}^2[O] & = \mathbb{E}_{\C} \mathbb{E}_{\mC} \left[O^2_{\mC} \right] - \left(\mathbb{E}_{\C} \mathbb{E}_{\mC} \left[O_{\mC} \right]\right)^2\\
    &= \underbrace{\mathbb{E}_{\C} \left[ \left(\mathbb{E}_{\mC}\left[O_{\mC} \right] \right)^2 \right] - \left( \mathbb{E}_{\C} \mathbb{E}_{\mC} \left[O_{\mC} \right] \right)^2}_{\text{circuit fluctuations}}
    +\mathbb{E}_{\C}\underbrace{ \left[\mathbb{E}_{\mC} \left[O^2_{\mC} \right] - \left( \mathbb{E}_{\mC} \left[O_{\mC} \right] \right)^2 \right]}_{\text{quantum fluctuations}}\\
    &= \sigma_{\C}^2[O] + \mathbb{E}_{\C}\left[ \sigma_{Q,\C}^2[O] \right]~, 
    \end{split}
    \end{equation}
\end{widetext}
where $\mathbb{E}_\C$ and $\mathbb{E}_{\mC}$ denote averages over circuits $\C$, and different trajectories $\mC$ given a circuit $\C$, respectively.
In going from the first to the second line in Eq.~\eqref{eq:fluct}, we have added and subtracted the quantity $\mathbb{E}_{\C} \left[\left( \mathbb{E}_{\mC} \left[O_{\mC} \right] \right)^2\right]$, where $O_{\mC}$ is for a single shot measurement of $O$ the measurement history $\mC$ given a circuit $\C$.
The first term, circuit fluctuations, pertains to the classical limit of choosing different circuits to evolve.
The second term, quantum fluctuations, is the variance of the outcome of measuring $O$ over many trials with the same circuit $\C$.
For the classical model, only the circuit fluctuations $\sigma_{\C}^2[O]$ are nonzero because the measurement outcome is deterministic, while for the quantum model, both circuit and quantum fluctuations can be nonzero.

Theoretically and, if we allow postselection, experimentally, the quantum fluctuations in Eq.~\eqref{eq:fluct} can be further decomposed into two parts:  
\begin{widetext}
\begin{equation}\label{eq:shots}
        \begin{split}
\sigma_{Q,\C}^2[O] &= \underbrace{\mathbb{E}_{\mC} \left[ \expval{O}_{\mC}^2 \right] - \left( \mathbb{E}_{\mC} \left[\expval{O}_{\mC} \right] \right)^2 }_{\text{trajectory fluctuations}} + \mathbb{E}_{\mC}\underbrace{ \left[\expval{O^2}_{\mC}-\expval{O}_{\mC}^2 \right] }_{\text{state fluctuations}} \\
&=\sigma_{t}^2[O] + \mathbb{E}_{\mC}[\sigma_{s,\mC}^2[O]],
        \end{split}
    \end{equation}
\end{widetext}
where $\expval{\dots}_{\bm{m}_c}$ denotes the quantum state expectation over the wave function $\ket{\psi_{\bm{m}_c}}$ of the system immediately before any measurement of $\dots$ is made. (Note that $\mathbb{E}_{\mC}\left[ \expval{O^n}_{\mC} \right] \equiv \mathbb{E}_{\mC} \left[ O^n_{\mC} \right]$ with $n=$1 and 2.)
The first term, trajectory fluctuations $\sigma_{t}^2[O]$, characterizes the uncertainty of different quantum trajectories $\mC$, coming from the randomness in the outcomes of the measurements that are made earlier than the measurement of $O$.
The second term, state fluctuations $\sigma_{s,\mC}^2[O]$, characterizes the quantum nature (superposition) of the pure state $\ket{\psi_{\mC}}$ immediately before the measurement of $O$ is made.  These two terms capture different aspects of the quantum fluctuations.  In the MIPT without feedback, the transition is seen only in properties of the state $\ket{\psi_{\mC}}$, although this is in entanglement properties rather than simple observables.

For our model with feedback, we will show that the trajectory fluctuations $\sigma_{t}^2[O]$ and state fluctuations $\sigma_{s,\mC}^2[O]$ both become critical at the CIPT, i.e., they obey single-parameter scaling. 
Whereas both parts of the quantum fluctuations are large for $p<p_{c}^{\text{CIPT}}$, they become small for $p>p_{c}^{\text{CIPT}}$, where {only small fluctuations around a classical trajectory remain.}
One open question about more general CIPT models is whether there exist models where the critical behavior differs between the trajectory fluctuations and the state fluctuations.

\subsubsection{Experimental tractability}\label{sec:experiments}

The separation of the observable fluctuations in Eq.~\eqref{eq:fluct} not only serves for a conceptual understanding of the quantum nature of the CIPT, but also provides a practical way to experimentally measure the quantum fluctuations, as this separation does not require post-selection.
The first term in Eq.~\eqref{eq:fluct}, the circuit fluctuations $\sigma_{\C}^2[O]$ [see Fig.~\ref{fig:schematic_fluct}(a)], can be experimentally sampled without post-selection. This is because the first contribution $\mathbb{E}_{\C} \left[ \left(\mathbb{E}_{\mC}\left[ {O}_{\mC} \right] \right)^2 \right]$ is obtainable by first fixing a specific circuit $\C$, and then randomly sampling the different measurement outcomes $\mC$ and the random shots at the final time following the Born rule given that $\C$ and $\mC$. 
The second contribution $\left( \mathbb{E}_{\C} \mathbb{E}_{\mC} \left[ {O}_{\mC} \right] \right)^2$ involves a simple average over all measurement outcomes and shots.

The second term in Eq.~\eqref{eq:fluct}, the quantum fluctuations $\sigma_{Q,\C}^2[O]$ can also be estimated without post-selection in experiments.
We can fix a specific circuit $\C$, randomly sample the trajectories $\mC$, and collapse the entire wave function at the final time to collect a single \textit{shot} of the observables $O_{\mC}$ and $O_{\mC}^2$. Repeating this process {for each sampled trajectory}, we can compute the variance given a specific circuit $\C$ to obtain the quantum fluctuations in Eq.~\eqref{eq:fluct}. 

However, separating the quantum fluctuations into the trajectory fluctuations $\sigma_{t}^2[O]$ [the first term in Eq.~\eqref{eq:shots} and also illustrated in Fig.~\ref{fig:schematic_fluct}(b)], and the state fluctuations $\sigma_{s,\mC}^2[O]$ [the second term in Eq.~\eqref{eq:shots} and also illustrated in Fig.~\ref{fig:schematic_fluct}(c)]-- in experiments seems to require postselection.
This is due to the measurement of the quantum average $\expval{O}^2_{\bm m_{\mathcal C}}$, which involves repeated realization of the same pure state $\ket{\psi_{\mC}}$ by post-selection.
Fortunately, in the Bernoulli map model with $\ket{0\dots0}$ as the control state, as we will show later, these two fluctuations share the same scaling dimension at the CIPT, and therefore, their combination shows the same critical behavior.
This also verified in Appendix~\ref{app:shots}, where we present a direct numerical simulation of the shot-to-shot fluctuation finding results consistent with those in Sec.~\ref{sec:traj_fluct} for the trajectory and state fluctuations, respectively.

\begin{figure*}[htbp]
    \centering
    \includegraphics[width=6.8in]{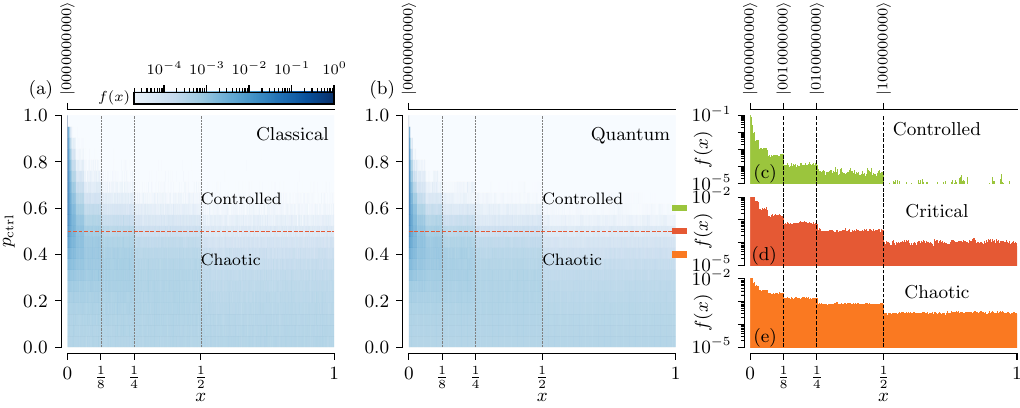}
    \caption{
    Ensemble-averaged bit string distribution $f(x)$ in Eq.~\eqref{eq:fidelity} as a function of $p_{\text{ctrl}}$ for (a) the classical model and (b) the quantum model, both with $L=10$.
    The bottom axis shows the number in base 10, and the top axis shows the corresponding bit strings in base 2.
    The false color shows $f(x)$ on a log scale (with an offset of $10^{-5}$ to avoid divergence at zero probability).
    (c-e) $f(x)$ for three typical $p_{\text{ctrl}}$ for the chaotic phase at $p_{\text{ctrl}}=0.4$ (orange), the critical point at $p_{\text{ctrl}}=0.5$ (red), and the controlled phase at $p_{\text{ctrl}}=0.6$ (green) for the quantum model.
    }
    \label{fig:bitstring_distribution}
\end{figure*}
\subsection{Summary of main results}\label{sec:main_results}

Having addressed these preliminaries, we now briefly summarize our main results regarding the quantum nature of the CIPT.

\begin{enumerate}
    \item At the level of ensemble-averaged wave function properties (e.g.~the ensemble-averaged distribution of the wave function over computational basis states), the quantum model looks similar to the classical model near the CIPT.
    \item The CIPT nevertheless manifests a phase transition in a quantum information-theoretic quantity: the ensemble-averaged $l_1$-coherence $\mathfrak{C}$ changes from an exponential growth as a function of system size $L$ in the chaotic phase to a saturating $L$-independent value in the controlled phase. At the critical point, the coherence growth remains exponential, indicating the critical wave function is quantum coherent.
    \item At the level of fluctuations in observables, the circuit-to-circuit fluctuations present in both the classical and quantum limits provide the dominant contribution and manifest the critical properties of the classical model.
    \item Although the trajectory-to-trajectory fluctuations, which appear only in the quantum limit, are subdominant, their contribution can be separated from the circuit-to-circuit fluctuations and measured experimentally. These intrinsic quantum fluctuations also become critical at the CIPT, 
    and develop a non-zero scaling dimension that
    governs the algebraic decay of these fluctuations at the critical point. 
    \item \HPadd{The exponential scaling of the quantum coherence and the critical behavior of both quantum fluctuation contributions (trajectory and state fluctuations) generalize to other adaptive monitored circuit models with an absorbing state transition.}
\end{enumerate}

These results provide a precise framework to demonstrate that the CIPT, which naively appears fully classical in nature, in fact,
manifests inherent quantum critical properties that were previously ``washed out'' (or hidden) by equally averaging over circuits and measurement outcomes.
We defer discussion of this point and further consequences to Sec.~\ref{sec:discussion}.

\section{Ensemble-Averaged wave function Properties}
\label{sec:mean}
Before discussing the different types of fluctuations, we first consider ensemble-averaged properties of the wave function, as it provides a direct connection to the classical limit.
We analyze the probability distribution of the bit-string-resolved wave function, ensemble averaged over different circuits and trajectories, without separating the quantum and classical contributions, and find that the quantum and classical results are essentially indistinguishable.
However, a distinction appears when we compute the ensemble-averaged $l_1$-coherence of the wave function to demonstrate that the system is quantum coherent at the CIPT. 
In the regime governed by quantum dynamics, the coherence is exponentially large (in the number of qubits $L$), and in the controlled phase, it saturates to an $L$-independent value.
At the CIPT critical point, it retains an exponential growth, indicating the average wave function is quantum coherent in the classical computational basis.
This implies that the CIPT in the quantum model has become more than the classical transition from which originates, and we therefore need to look at its quantum effects in a precise manner, which is the focus of the section following this one.

\subsection{Ensemble-averaged bit string distribution}\label{sec:bitstring}
To explicitly visualize how the control map is imprinted upon the quantum state, we plot the ensemble-averaged probability distribution over bit strings as a function of the control rate $p_{\text{ctrl}}$ for both the classical [Fig.~\ref{fig:bitstring_distribution}(a)] and quantum [Fig.~\ref{fig:bitstring_distribution}(b)] Bernoulli circuit models for system size $L=10$.
Here, the probability of finding the state in a specific bit string $x$ is defined as
\begin{equation}\label{eq:fidelity}
    f(x)=\mathbb{E}_{\bm{M}}  \left[ \abs{\braket{x}{\psi_{\bm{M}}}}^2 \right],
\end{equation}
where $\ket{\psi_{\bm{M}}}$ is the steady state of the system for a specific sample $\bm{M}$ including a fixed circuit realization $\C$ and measurement history $m_{\C}$.
In the classical model, the system is in a simple product state at all times.

In Fig.~\ref{fig:bitstring_distribution}(a,b), we notice a striking visual resemblance between the classical and quantum models---as the control rate $p_{\text{ctrl}}$ increases from 0, the distribution of the bit string changes from a uniform distribution over all bit strings (corresponding to an ergodic chaotic phase) to a distribution localized at the fixed point of the polarized state $\ket{00\dots0}$. 

To understand the structure in the computational basis in more detail we turn to  Figs.~\ref{fig:bitstring_distribution}(c-e), where we present bit string distributions $f(x)$ for three typical values of $p_{\text{ctrl}}$ corresponding to the chaotic phase at $p_{\text{ctrl}}=0.4$ (orange), the critical point at $p_{\text{ctrl}}=0.5$ (red), and the controlled phase at $p_{\text{ctrl}}=0.6$ (green), respectively. 
We see for $p_{\mathrm{ctrl}}<p_{\mathrm{ctrl}}^{c}$ that $f(x)$ is almost uniformly distributed across all the basis states, namely $f(x)\sim dx = 2^{-L} $, where $dx$ is the measure coming from the discretization of the interval $[0,1)$.
For $p_{\mathrm{ctrl}}>p_{\mathrm{ctrl}}^{c}$, we find $f(x)\sim \delta_{x} dx$ up to a tail that goes to zero in the thermodynamic limit (see Appendix~\ref{app:distribution} for more details on the scaling of the distribution with system size $L$), demonstrating that the wave function ``sticks'' to the fixed point in the controlled phase. 
At the CIPT $p_{\mathrm{ctrl}}=p_{\mathrm{ctrl}}^{c}$, the bit string distribution takes the power law form $f(x)\sim L^{-1} dx$.

An interesting feature is the staircase pattern in the distribution of bit strings, $f(x)$, at a given $p_{\text{ctrl}}$, where abrupt changes all happen at bit strings that are inverse of the power of 2, and probability density within each interval remains almost uniform.
The positions of abrupt changes motivate us to define the position of FDW measured as the first (leftmost) ``1" from the right edge of the bit string $x$. 
For example, $\ket{0000000010}$ and $\ket{0000000011}$ have $k=2$, and the polarized state $\ket{0000000000}$ has $k=0$.
Formally, we can define the FDW position for $x\in[0,1)$ as
\begin{equation}\label{eq:k}
    k=\text{FDW}(x)=\begin{cases}
        \log_2(\lfloor x2^L \rfloor)+1, \quad & 0<x<1\\
        0, \quad & x=0
    \end{cases}
\end{equation}
where $\lfloor\dots\rfloor$ is the floor function. 

Therefore, the state of the system can be effectively described by the FDW location $k$ instead of tracking the entire wave function, hence providing an effective metric to monitor its dynamics. (see Appendix~\ref{app:distribution} for more quantitative analysis of the FDW distribution.)

\subsection{Quantum coherence}\label{sec:coherence}

\begin{figure}[htbp]
    \centering
    \includegraphics[width=3.4in]{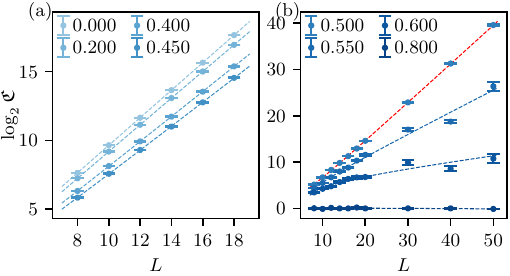}
    \caption{
        Quantum coherence $\mathfrak{C}$  [Eq.~\eqref{eq:coherence}] as a function of $p_{\text{ctrl}}$ in (a) chaotic phase showing exponential growth with system size $L$ using state vector evolution, and (b) the critical \HPadd{(red dashed line)} and controlled phase showing sub-exponential growth using matrix product states.
        The dashed line is the fitting of the coherence following Eq.~\eqref{eq:coherence_fiting}. The fitted parameters are shown in Fig.~\ref{fig:coherence_fitting}.
        }
    \label{fig:coherence_L}
\end{figure}
\begin{figure*}[htbp]
    \centering
    \includegraphics[width=6.8in]{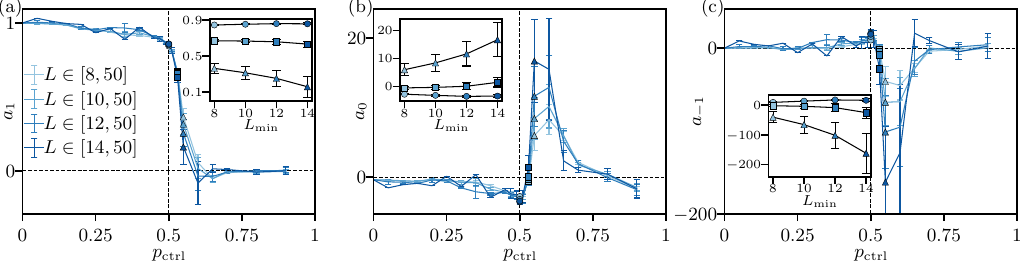}
    \caption{
        The fitted parameters (a) $a_1$, (b) $a_0$, and (c) $a_{-1}$ appearing in the series expansion of the coherence $\mathfrak{C}$ posited in Eq.~\eqref{eq:coherence_fiting}. Above the CIPT, the different curves correspond to different system size ranges $[L_{\rm min},50]$ with varying $L_{\rm min}$ as shown in the legend.
        Below the CIPT, system sizes $L\in[8,18]$ are considered.
        The insets show the trends of the fitted parameters as a function of $L_{\text{min}}$ for $p_{\text{ctrl}}=0.5$ (circles),   $p_{\text{ctrl}}=0.53$ (squares), and $p_{\text{ctrl}}=0.55$ (triangles).
        }
    \label{fig:coherence_fitting}
\end{figure*}

In the previous section, we found that the ensemble-averaged FDW distribution is the same for both the classical and quantum models (see Appendix~\ref{app:distribution} for a quantitative analysis of the two distributions), which does not inform us about the quantum nature of the CIPT. 
However, we can probe quantum effects by directly looking at the essence of a \textit{quantum} state, i.e., its coherence due to superposition.

The coherence measure we adopt here is the ensemble-averaged $l_1$-quantum coherence~\cite{baumgratz2014quantifying}
\begin{equation}\label{eq:coherence}
    \mathfrak{C}=\mathbb{E}_{\bm{M}}\left[ \sum_{x\neq x'} \abs{(\rho_{\bm{M}})_{x,x'}} \right],
\end{equation}
where $\rho_{\bm{M}}$ is the density matrix for a specific sample $\bm{M}$, and ${(\rho_{\bm{M}})_{x,x'}}$ is the off-diagonal element of the density matrix corresponding to $\ketbra{x}{x'} =\bigotimes_{i=1}^L\ketbra{b_i}{b_i'}$.  
The summation runs over all the off-diagonal elements of the density matrix, which originate from multiple superposition components within the quantum state.
Consequently, the $l_1$-coherence is manifestly zero in the classical model.
Note that we choose to focus on the \textit{average coherence} of the state rather than the coherence of the average state as, in the latter case, the sample average washes out the off-diagonal density-matrix elements, leading to zero coherence.
This is similar to the reason why ensemble-averaged entanglement measures are considered in studies of MIPTs.

In our numerical simulations of the coherence, we use both state-vector and MPS evolution algorithms.
We use state-vector evolution to compute the coherence throughout the phase diagram (including the volume-law phase) up to $L=18$.
Within the area-law phase, which includes the CIPT, the object of our study, we use MPS to simulate system sizes beyond $L=20$.
Here, the nontrivial part is to compute the element-wise absolute value of the MPS wave function without fully contracting the internal legs of the MPS, which would lead to an exponential memory cost.
We, therefore, use tensor cross interpolation~\cite{oseledets2010ttcross,dolgov2020parallel,ritter2024quantics,fernandez2024learning} (TCI) to learn the mapping from the physical index of an MPS to the norm of the corresponding wave function amplitude:
\begin{equation}\label{eq:TCI}
    \mathfrak{f}:(b_1,b_2,\dots, b_L) \in \mathbb{Z}_2^{\otimes{L}} \mapsto \abs{\braket{b_1b_2\dots b_L}{\psi_{\bm{M}}}}\in \mathbb{R},
\end{equation} 
where $\ket{b_1b_2\dots b_L}$ corresponds to the computational basis state $\ket{x}$ with $x\in[0,1)$. 
With TCI, we can directly obtain each element in the density matrix---in complexity $O(\chi^4)$ where $\chi$ is the maximal bond dimension of the MPS---without fully contracting the MPS. 
We choose a sufficiently large maximal bond dimension to capture the wave function within the area-law phase. More details on the TCI method are provided in Appendix~\ref{app:TCI}.

We start with an initial product state with FDW location $k=1$ and evolve the system to the steady state, which is reached after $2L^2$ time steps.
In Fig.~\ref{fig:coherence_L}, we plot the steady-state coherence $\log_2\mathfrak{C}$ as a function of system size $L$ for different $p_{\text{ctrl}}$ using state-vector evolution for the chaotic phase in Fig.~\ref{fig:coherence_L}(a) and MPS for the critical and controlled phases in Fig.~\ref{fig:coherence_L}(b).
In the chaotic phase [Fig.~\ref{fig:coherence_L}(a)], we find that the coherence grows exponentially with $L$, namely $\mathfrak{C} \sim 2^{a_1(p)L}$.
At $p_{\text{ctrl}}=0$, we analytically find $\mathfrak{C}=\frac{\pi}{4}2^L$ as $L\rightarrow\infty$ from the coherence of a Haar random state (see Appendix~\ref{app:coherence_Haar}), which is the maximal possible coherence.
For a finite $p_{\text{ctrl}}$ in the chaotic phase, the coherence continues to grow exponentially, however, with $0<a_1(p)<1$.
Surprisingly, we find that at the critical point, the coherence remains exponential with $a_1(p_c)\approx 0.85$.
In contrast, for $p>p_c$, we find that $\mathfrak{C}$ saturates to a system size independent value. 
This motivates the hypothesis that the coherence can be parameterized as
$\mathfrak{C}\sim 2^{h(p,L)}$, with $h(p,L)$ becoming non-analytic at the control transition.
To test this, we apply an empirical ``proof by contradiction'' by assuming $h(p,L)$ is analytic in $p$, so that an asymptotic expansion of the form
\begin{equation}\label{eq:coherence_fiting}
    \log_2\mathfrak{C} \sim {a_1(p) L+a_0(p)+a_{-1}(p)L^{-1} +O(L^{-2})}
\end{equation}
holds.
We will show that the coefficients of this expansion become singular near the critical point, demonstrating that $h(p,L)$ becomes non-analytic at this dynamical quantum phase transition.

We analyze the fitted parameters $a_{1}$, $a_0$, and $a_{-1}$ in Fig.~\ref{fig:coherence_fitting} by gradually excluding smaller system sizes from the fitting.
In Fig.~\ref{fig:coherence_fitting}(a), we find that $a_1$, which parameterizes the exponential growth of the coherence, remains almost constant throughout the chaotic phase up to the CIPT, where it jumps to zero. 
As we continue to exclude smaller system sizes, the jump becomes more abrupt, as shown in the inset.

In Fig.~\ref{fig:coherence_fitting}(b), we study $a_0$, which parameterizes the system-size independent contribution. Starting with $a_0\approx-0.34\approx\log_2(\pi/4)$ at zero control rate, we find that $a_0$ decreases monotonically with $p_{\text{ctrl}}$ and reaches its minimum at the critical point, which is consistent with the decreasing growth exponent in Fig.~\ref{fig:coherence_L}(a). 
In the controlled phase at $p_{\text{ctrl}}>0.6$, we find that $a_0$ also decreases monotonically but from a large positive value.
Here, $a_0$ corresponds to the saturation coherence at large $L$, which is consistent with the decreasing saturation coherence for larger $p_{\text{ctrl}}$ in Fig.~\ref{fig:coherence_L}(b).
The growth from the minimum value to the large positive value between $p_{\text{ctrl}}=0.5$ and $p_{\text{ctrl}}=0.55$ is a  finite-size effect: The data in the inset show that the peak of $a_0$ in the controlled phase near $p_{\text{ctrl}}=0.55$ becomes sharper as we exclude smaller system sizes from the fitting as shown in the inset.

Finally, in Fig.~\ref{fig:coherence_fitting}(c), we show data for $a_{-1}$, which parameterizes the subleading correction of order $L^{-1}$. We notice that it remains zero in both the chaotic and controlled phases but diverges at the critical point, consistent with the function $h(p,L)$ becoming non-analytic at the transition in the thermodynamic limit $(L\rightarrow\infty)$. It is striking that the coherence remains exponentially growing at the transition, demonstrating that the critical wave function is inherently quantum mechanical.
\HPadd{This exponential scaling of the coherence at the critical point generalizes to other adaptive monitored circuit models with an absorbing state transition, as we will show in Sec.~\ref{sec:generalization}.}

\section{Classical vs. quantum fluctuations of observables}\label{sec:fluctuation}
The critical behavior of the quantum coherence already demonstrates the quantum nature of the CIPT in the quantum model.
However, quantum coherence is challenging to measure experimentally as, like the entanglement entropy, one must compute the coherence of each sample before averaging over samples.
It is therefore preferable to try to access the quantum nature of the CIPT with quantities that do not require post-selection.
In Sec.~\ref{sec:mean}, we showed that the ensemble-averaged bit string distribution is dominated by classical behavior.
Nevertheless, as discussed in Sec.~\ref{sec:separating}, considering the sample-to-sample fluctuations of an observable allows us to separate out quantum and classical fluctuations.
In the following results, we will demonstrate that the quantum nature of the CIPT can be observed in the quantum contribution to these fluctuations.

\begin{figure}[htbp]
    \centering
    \includegraphics[width=3.4in]{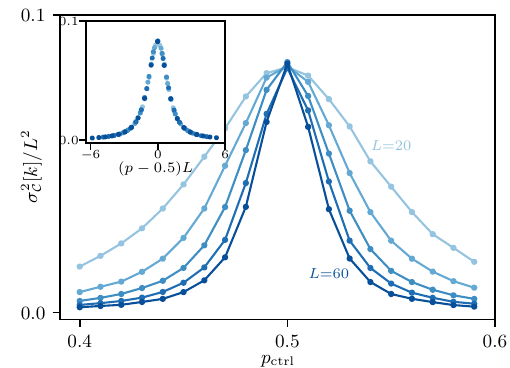}
    \caption{
    Steady-state circuit-to-circuit fluctuations at $p_{\text{ctrl}}=0.5$ for different system sizes $L$.
    The inset shows finite-size-scaling collapse following Eq.~\eqref{eq:fss_sigma_c} with $\nu=1.007(2) $  and $p_c^{\text{CIPT}}=0.500(1)$.
    }
    \label{fig:cl_fluct_FSS}
\end{figure}
\subsection{Classical circuit-to-circuit fluctuations}\label{sec:classical_fluctuation}
We first consider the dominant contribution to the unresolved fluctuations in the quantum model, namely the classical circuit-to-circuit fluctuations
\begin{equation}\label{eq:cl_fluct}
    \sigma_{\C}^2[k]=\mathbb{E}_{\C} \left[ \left(\mathbb{E}_{\mC}\left[ {k}_{\mC} \right] \right)^2 \right] - \left( \mathbb{E}_{\C} \mathbb{E}_{\mC} \left[ {k}_{\mC} \right] \right)^2,
\end{equation}
{where $\mathbb{E}_{\mC}\left[ {k}_{\mC} \right]$ is the trajectory-averaged steady-state FDW expected value given a specific circuit $\C$.}

To probe the classical fluctuations, we choose an initial state with the FDW at $k=L/2$, and allow the circuit to reach the steady state (the dynamics will be discussed in a sequel~\cite{panappear}). 
The classical nature of the circuit-to-circuit fluctuation can be seen in the steady-state value $\sigma_{\C}^2[k]$ as a function of $p_{\rm ctrl}$ for different system sizes $L$, as shown in Fig.~\ref{fig:cl_fluct_FSS}.
Here, we find that the steady-state fluctuations peak at the critical point, with the peak value scaling as $O(L^2)$, consistent with the dynamical exponent of $z=2$ in the unbiased random walk.
With the finite-size scaling ansatz
\begin{equation}\label{eq:fss_sigma_c}
    \sigma_{\C}^2[k]/L^2 \sim f_{\C}\left( \left(p_{\text{ctrl}}-p_c^{\text{CIPT}}\right)L^{1/\nu} \right),
\end{equation}
we find data collapse as shown in the inset of Fig.~\ref{fig:cl_fluct_FSS} with the estimated critical exponents $\nu=1.007(2)$ and $p_c^{\text{CIPT}}=0.500(1)$ consistent with the random walk universality reported in Ref.~\cite{iadecola2023measurement}.

\begin{figure*}[htbp]
    \centering
    \includegraphics[width=6.8in]{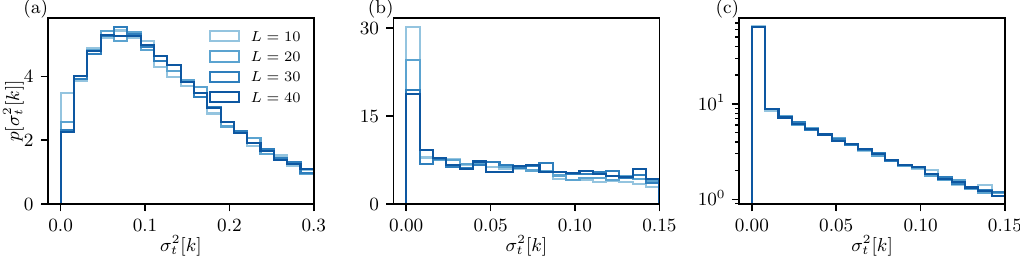}
    \caption{
    Probability distribution over all circuits $\C$ of the trajectory fluctuations $\sigma_{t}^2[k]$ for steady states at (a) $p_{\text{ctrl}}=0.4$, (b) $p_{\text{ctrl}}=0.5$, and (c) $p_{\text{ctrl}}=0.6$ across several system sizes.
    A clear distinction between each phase is observed, with a vanishing weight at $\sigma_{t}^2[k]=0$ in the chaotic phase, which develops into a non-zero contribution in the control phase, demonstrating the dynamics are dominated by a single classical trajectory. At the transition, we see that the classical contribution goes to zero in the thermodynamic limit.
    }
    \label{fig:traj_fluct}
\end{figure*}
\begin{figure}[htbp]
    \centering
    \includegraphics[width=3.4in]{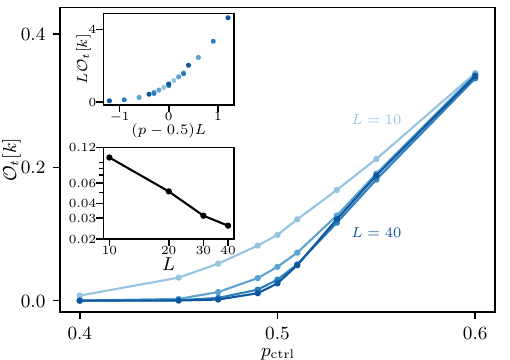}
    \caption{
    Order parameter $\mathcal{O}_{t}[k]$ for the trajectory fluctuation of the first domain wall $k$ [Eq.~\eqref{eq:op}] for different system sizes $L$.
    The full ensemble includes 500 different circuits $\C$, and 500 different trajectories $\mC$ per circuit.
    {Upper inset: Finite-size scaling of the order parameter following Eq.~\eqref{eq:op} with $\nu=1.00(5)$, $p_c^{\text{CIPT}}=0.50(2)$, and $\beta_t=0.98(9)$.}
    {Lower inset: Algebraic decay of the order parameter as a function of $L$ at the critical point.}
    }
    \label{fig:traj_fluct_FSS}
\end{figure}
\subsection{Fluctuations across quantum trajectories}\label{sec:traj_fluct}    
The steady-state trajectory fluctuation $\sigma_{t}^2[k]$ in Eq.~\eqref{eq:observable_fluctuation}, which measures fluctuations across different quantum trajectories for a fixed circuit realization, is unique to the quantum setting as it is only nonzero when the measurement outcomes are random:
\begin{equation}\label{eq:traj_fluct}
        \sigma_{t}^2[k] = \mathbb{E}_{\mC} \left[ \expval{k}_{\mC}^2 \right] - \left( \mathbb{E}_{\mC} \left[ {k}_{\mC} \right] \right)^2,
\end{equation}
where $\expval{k}_{\mC}=\expval{k}{\psi_{\mC}}$ is the steady state FDW expected value for a specific quantum trajectory $\mC$ given a specific circuit $\C$. 
It is subdominant [$\sigma_{t}^2[k]\sim O(1)$] compared to the classical circuit-to-circuit fluctuations [$\sigma_{\C}^2[k]\sim O(L^2)$], so its contribution is not visible in the full average over samples. However, as discussed in Sec.~\ref{sec:separating}, it can be measured independently in experiments.
We also examine the trajectory fluctuation using the magnetization density $M_z$ in Appendix~\ref{app:fluctuation_O}, finding results consistent with those for the FDW.

As an illustrative example, we first visualize the temporal spreading of the trajectories of the averaged FDW $k(t)=\mathbb{E}_{\mC}\left[ \expval{k}{\psi(t)_{\mC}} \right]$ in Fig.~\ref{fig:schematic}(e) starting from $\expval{k(0)}=1$ for the three representative control rates $p_{\text{ctrl}}=0.4$, $0.5$, and $0.6$ in the quantum model.
Within each panel, we plot a collection of quantum trajectories given the same circuit realization $\C$ (i.e., the same realization of the unitary scrambler $U$ in the Bernoulli map and the same sequence of chaotic and control time steps). 
The visible fluctuations come from the different measurement outcomes $\mC$.
In the chaotic phase (left panel), the trajectories start to diverge immediately from the initial state at $k=1$ and quickly saturate to the left end of the chain at $k=L$. 
The large variance over different trajectories reflects the nature of the quantum dynamics due to the chaos in the Bernoulli map.
However, in the controlled phase (right panel), the trajectories remain close to the fixed point, overlapping with each other, and the spreading of the trajectories only happens when the chaotic map is applied multiple times in a row (which is a rare event inside the controlled phase). 
This demonstrates the scenario where the quantum dynamics with a large spreading of trajectories is reduced to an effectively classical dynamics with a common trajectory.
At the critical point in the middle panel, we see critical behavior where the FDW undergoes an unbiased random walk, and the spreading of the trajectories remains $O(1)$.

Figure~\ref{fig:schematic}(e) provides an intuitive picture of the trajectory fluctuations. 
To quantify these fluctuations, we plot in Figs.~\ref{fig:traj_fluct}(a-c) the probability distribution of $\sigma_{t}^2[k]$ in the steady state for the three representative values $p_{\text{ctrl}}=0.4$, 0.5, and 0.6 over different circuit realizations $\C$.
\footnote{We evolve the circuit up to $t=2L^2$, and confirm that the dynamics reaches a steady state for $t>L^2$. 
Therefore, all states within $t\in[L^2,2L^2]$ can be treated as separate samples for constructing the histogram in Fig.~\ref{fig:traj_fluct} to enhance the smoothness of the statistics with minimal additional computational cost. We also confirm the qualitatively same results using only the data at final time step $t=2L^2$.}
We find that in the chaotic phase [Fig.~\ref{fig:traj_fluct}(a)], most of the trajectory fluctuations are finite (around $0.1$) and it is unlikely to have zero fluctuations in the thermodynamic limit---the probability of zero fluctuations is exponentially small in system size.
At the critical point [Fig.~\ref{fig:traj_fluct}(b)], the probability of zero fluctuations instead decays algebraically as a function of system size.
In the controlled phase [Fig.~\ref{fig:traj_fluct}(c)], most of the trajectory fluctuations are zero regardless of system size with an exponential tail of finite fluctuations, indicating classical dynamics that converge around a single trajectory with deterministic measurement outcomes.

This change in the distribution of the trajectory fluctuations $\sigma_{t}^2[k]$ as a function of $p_{\text{ctrl}}$ motivates us to define an order parameter $\mathcal{O}_{t}[k]$ as the probability of zero fluctuations,
\begin{equation}\label{eq:op}
    \mathcal{O}_{t}[k] = P[\sigma_{t}^2[k]=0]\approx \int_0^\epsilon p[\sigma_{t}^2[k]] d \sigma_{t}^2[k],
\end{equation}
where $\epsilon=10^{-5}$ is a small numerical cutoff.
This order parameter is shown in Fig.~\ref{fig:traj_fluct_FSS} as a function of $p_{\rm ctrl}$ for different system sizes $L$.
It vanishes in the chaotic phase, decays algebraically at the critical point, and saturates to a finite value in the controlled phase.

The algebraic decay of the order parameter $\mathcal{O}_{t}[k]$ at the critical point exposes another {scaling dimension} defined as 
\begin{equation}
    \mathcal{O}_{t}[k]\sim \begin{cases}
        \left( p_{\text{ctrl}}-p_c^{\text{CIPT}} \right)^{\beta_t} ,& p_{\text{ctrl}}>p_c^{\text{CIPT}} ,\\
        0 , & p_{\text{ctrl}}\le p_c^{\text{CIPT}}
    \end{cases}
\end{equation}
in the $t\rightarrow\infty$ and $L\rightarrow\infty$ limit, characterizing the onset of classicality.
{Here, we denote the scaling dimension by $\beta$ to make an analogy to the order-parameter critical exponent in the conventional context of continuous phase transitions.}
In the classical model, $\beta_t$ does not exist since the trajectory fluctuations $\sigma_{t}^2[k]$ are manifestly zero, and therefore $\mathcal{O}_{t}[k]$ will be always 1.
In the quantum model, to extract the scaling dimension $\beta_t$, we perform finite-size scaling of the order parameter $\mathcal{O}_{t}[k]$ following the scaling form 
\begin{equation}\label{eq:fss}
    \mathcal{O}_{t}[k] \sim L^{-\beta_t/\nu}f_t\left( \left(p_{\text{ctrl}}-p_c^{\text{CIPT}}\right)L^{1/\nu} \right).
\end{equation}
In the inset of Fig.~\ref{fig:traj_fluct_FSS}, we estimate the {new scaling dimension} as $\beta_t=0.98(8)$, along with $\nu=0.99(3)$, $p_c^{\text{CIPT}}=0.50(2)$ consistent with the classical results.
This unveils another inherently quantum aspect of the CIPT which was previously overlooked due to the subleading nature of the quantum contribution to the sample-to-sample fluctuations.
These universal quantum fluctuations should be generic to other adaptive quantum dynamics as well.

\subsection{Fluctuations across quantum states}\label{sec:state_fluct}
\begin{figure}[htbp]
    \centering
    \includegraphics[width=3.4in]{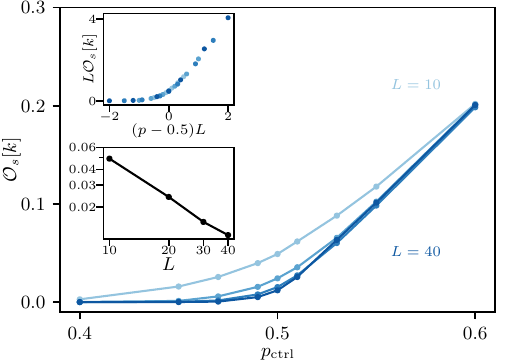}
    \caption{
    Order parameter $\mathcal{O}_s[k]$ for the state fluctuation of the first domain wall $k$ for different system sizes $L$ as a function of $p_{\rm ctrl}$. 
    Upper inset: Finite-size scaling collapse with $\nu=1.00(3)$, $p_c^{\text{CIPT}}=0.500(3)$, and $\beta_{s}=1.00(1)$.
    Lower inset: Algebraic decay of the order parameter as a function of $L$ at the critical point.
    }
    \label{fig:state_fluct_FSS}
\end{figure}
Having explored trajectory fluctuations in detail we now turn to a state fluctuation $\sigma_{s,\mC}^2[k]$ in Eq.~\eqref{eq:fluct}. 
Similarly, we study the FDW location $k$ as
\begin{equation}\label{eq:state_fluct}
    \sigma_{s,\mC}^2[k]=  \expval{k^2}_{\mC} - \expval{k}_{\mC}^2 
\end{equation}
where $\expval{k^2}_{\mC}=\expval{k^2}{\psi_{\mC}}$ is the second moment of $k$ for a specific quantum trajectory ${\mC}$ and a fixed circuit $\C$.
This quantity intuitively estimates the average spreading of the FDW wave packet within a single quantum trajectory.
For the classical model, Eq.~\eqref{eq:state_fluct} is also by definition zero because the ``wave function'' $\ket{\psi_{\bm{m}_\C}}$ is a product state with a sharp FDW location, as opposed to a quantum wave packet.
For the quantum model, the Bernoulli map generates quantum coherence as it scrambles a product state into a superposition of product states.

Similar to the trajectory fluctuations in Eq.~\eqref{eq:op}, the distribution of the state fluctuations at the steady state can also be used to unveil the quantum nature of the CIPT.
Here, we can define the probability of zero fluctuations in $\sigma_{s,\mC}^2[k]$ for different circuits $\C$ and trajectories $\mC$ as an order parameter $\mathcal{O}_s[k]$ similar to Eq.~\eqref{eq:op}.
The order parameter $\mathcal{O}_s[k]$ also changes from zero in the chaotic phase to a finite value in the controlled phase with an algebraic decay at the critical point, as shown in the lower inset in Fig.~\ref{fig:state_fluct_FSS}, similar to Eq.~\eqref{eq:op}.
To extract the corresponding {scaling dimension} $\beta_{s}$, we plot in Fig.~\ref{fig:state_fluct_FSS} the order parameter $\mathcal{O}_s[k]$ as a function of $p_{\text{ctrl}}$ for $L=10$ to $L=40$.
The upper inset in Fig.~\ref{fig:state_fluct_FSS} shows the finite-size scaling of the order parameter $\mathcal{O}_s[k]$ following the scaling form in Eq.~\eqref{eq:fss} to find the scaling dimension $\beta_{s}=0.99(3)$, with critical exponent $\nu=0.99(1)$, and critical measurement rate $p_c^{\text{CIPT}}=0.500(1)$, which is consistent with the critical exponents extracted from the trajectory fluctuation in Sec.~\ref{sec:traj_fluct}.

{Although both the trajectory and state fluctuations in the Bernoulli map model exhibit the same critical behavior at the CIPT, we notice that they are quantitatively different.
In the controlled phase, the trajectory fluctuations (Fig.~\ref{fig:traj_fluct_FSS}) have a larger order parameter than the state fluctuations in Fig.~\ref{fig:state_fluct_FSS}, indicating that the states with zero state fluctuations are a subset of those with zero trajectory fluctuations, given the ferromagnetic state as the dark state.
This suggests that the state fluctuations being zero is a stronger condition than the trajectory fluctuations being zero. 
{For example, when the states across all trajectories are reset to the fixed point $\ket{0\dots0}$ (e.g., after many consecutive applications of the control map), a \textit{fixed} Bernoulli map in the next time step drives all states to the same state $z_0\ket{0\dots00}+z_1\ket{0\dots01}+z_2\ket{0\dots10}+z_3\ket{0\dots11}$, leading to a zero trajectory fluctuation, while finite state fluctuation due to the coherence.}

The opposite scenario with finite trajectory fluctuations but zero state fluctuations can also exist. For example, in the conventional Haar-random MIPT bricklayer model without any feedback, the trajectory fluctuations are always finite regardless of the measurement rate, while the state fluctuations are zero beyond the connectivity transition. 
Adaptive circuits with other feedback operations targeting other dark states can lead to different qualitative behaviors for the two types of quantum fluctuations, which is an interesting direction for future research.} 

\section{Generalization to other absorbing-state transitions}\label{sec:generalization}
In previous sections, we have focused on the Bernoulli map as a concrete example to illustrate the quantum nature of the CIPT in terms of both the quantum coherence and the quantum fluctuations of observables. 
In fact, this framework is model-independent and can be generically applied to any quantum circuit showing absorbing-state transitions.
Therefore, in the following sections, we provide two other examples of absorbing-state transitions with exact fixed points:
a Haar random adaptive circuit~\cite{odea2024entanglement}, and a bricklayer Clifford random adaptive monitored circuit~\cite{sierant2023controlling}.

\subsection{Haar random adaptive circuit}
The first generalization is to the Haar random adaptive circuit in Ref.~\cite{odea2024entanglement}, where the unitary process is built from two-qubit gates of the form $\mathcal{U}=\mathds{1}\oplus \mathfrak{u}(3)$ up to a $\mathfrak{u}(1)$ global phase.
Here, $\mathds{1}$ acts trivially on the basis state $\ket{00}$, and therefore preserves the ferromagnetic absorbing state $\ket{0}^{\otimes L}$ in an $L$-site 1D qubit chain; $\mathfrak{u}(3)$ is a Haar-random unitary that scrambles the other three basis states $\ket{01}$, $\ket{10}$, and $\ket{11}$. 
The feedback consists of local reset operations in which a qubit is measured in the computational basis, and flipped only if the measurement outcome is $\ket{1}$ (i.e., we choose to always correct with $p_f=1$ using the notation in Ref.~\cite{odea2024entanglement}); otherwise, the qubit remains in $\ket{0}$.
The circuit follows a bricklayer structure, with two-qubit random unitaries applied to even and odd pairs of qubits alternately, and interspersed with single-qubit measurements followed by feedback with probability $p_m^A$. 
The absorbing-state transition happens at $p_{m,c}^A=0.09085(5)$ with a dynamical exponent of $z^A=1.6(1)$ and a critical exponent of $\nu^A=1.07$~\cite{odea2024entanglement}. 
The constrast between this model and the Bernoulli map lies in the nature of the unitary dynamics: the ferromagnetic state is a true absorbing state in the the Haar random adaptive circuit, while it is only an approximate absorbing state in the Bernoulli map. 

The order parameter to probe the absorbing-state transition can similarly be chosen as the defect density 
\begin{equation}\label{eq:n_d}
    n_d=\frac{1}{L}\sum_{i=1}^{L}  \frac{1-Z_i}{2} ,
\end{equation}
which is related to the magnetization density in Eq.~\eqref{eq:magnetization} by a shift and rescaling.

\subsubsection{Quantum coherence}\label{app:absorbing}
\begin{figure}[htbp]
    \centering
    \includegraphics[width=3.4in]{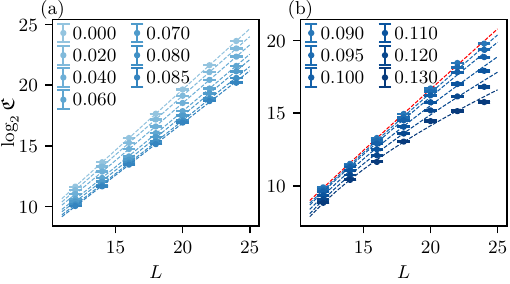}
    \caption{
    Quantum coherence $\mathfrak{C}$ [Eq.~\eqref{eq:coherence}] for the Haar random adaptive circuit absorbing-state transition model defined in Ref.~\onlinecite{odea2024entanglement} as a function of $p_m^A$ in (a) non-absorbing phase showing exponential growth with system size $L$ using state vector evolution, and (b) the critical (red dashed line) and absorbing phase showing sub-exponential growth using matrix product states. 
    The evolution time step is set to be $0.12L^{z^A}$. 
    The dashed line is the fitting of the coherence following Eq.~\eqref{eq:coherence_fiting}. The fitted parameters are shown in Fig.~\ref{fig:coherence_fitting_APT}. 
    }
    \label{fig:coherence_L_APT}
\end{figure}

We first study the $l_1$-quantum coherence in Eq.~\eqref{eq:coherence} for the absorbing-state transition model, as shown in Fig.~\ref{fig:coherence_L_APT}(a) for the non-absorbing phase ($p_m^A<p_{m,c}^A$) using state-vector evolution, and in Fig.~\ref{fig:coherence_L_APT}(b) for the critical point and absorbing phase ($p_m^A\gtrsim p_{m,c}^A$).
We find qualitatively similar behavior: the coherence grows exponentially with system size $L$ in the non-absorbing phase, and sub-exponentially in the absorbing phase.

\begin{figure*}[htbp]
    \centering
    \includegraphics[width=6.8in]{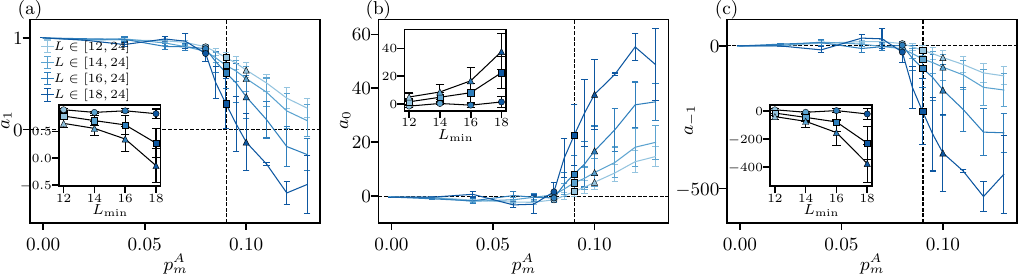}
    \caption{
        The fitted parameters (a) $a_1$, (b) $a_0$, and (c) $a_{-1}$ appearing in the series expansion of the coherence $\mathfrak{C}$ posited in Eq.~\eqref{eq:coherence_fiting}. 
        The different curves correspond to different system size ranges $[L_{\rm min},24]$ with varying $L_{\rm min}$ as shown in the legend.        
        The insets show the trends of the fitted parameters as a function of $L_{\text{min}}$ for $p_m^A=0.08$ (circles),   $p_m^A=0.09$ (squares), and $p_m^A=0.1$ (triangles).
        }
    \label{fig:coherence_fitting_APT}
\end{figure*}

Using the same ``proof by contradiction'' strategy as in Sec.~\ref{sec:coherence}, we fit the log of the average coherence using the series expansion in Eq.~\eqref{eq:coherence_fiting}, and analyze the fitted parameters $a_1$, $a_0$, and $a_{-1}$ in Fig.~\ref{fig:coherence_fitting_APT}.
We find that $a_1$ remains almost $O(1)$ in the non-absorbing phase and jumps to zero beyond the critical point as indicated by the vertical dashed line, and the jump becomes more abrupt as we exclude smaller system sizes from the fitting.
In Fig.~\ref{fig:coherence_fitting_APT}(b), we find that $a_0\approx -0.3$, which is slightly smaller than the true Haar random value of $\log_2(\pi/4)$, which is due to the nature of the unitary used in here being less random than the full two-qubit random unitary in $\mathfrak{u}(4)$.
Finally, in Fig.~\ref{fig:coherence_fitting_APT}(c), we find that $a_{-1}$ diverges after the critical point, consistent with the scaling behaviors described previously in Sec.~\ref{sec:coherence}.


\subsubsection{Quantum fluctuation}
Given the order parameter $n_d$ in Eq.~\eqref{eq:n_d}, we can define the trajectory fluctuation as 
\begin{equation}\label{eq:traj_fluct_APT}
    \sigma_{t}^2[n_d] = \mathbb{E}_{\mC} \left[ \expval{n_d}_{\mC}^2 \right] - \left( \mathbb{E}_{\mC} \left[ \left({{n_d}}\right)_{\mC} \right] \right)^2,
\end{equation}
similar to Eq.~\eqref{eq:traj_fluct} but with a different observable, 
and the state fluctuation as
\begin{equation}\label{eq:state_fluct_APT}
    \sigma_{s,\mC}^2[n_d]=  \expval{n_d^2}_{\mC} - \expval{n_d}_{\mC}^2 .
\end{equation}
similar to Eq.~\eqref{eq:state_fluct}.
Here, each circuit $\C$ is defined as trajectories with the same realization of the random unitaries $\mathcal{U}$ and the position of the measurements, while each trajectory $\mC$ is associated with a specific sequence of measurement outcomes in the circuit $\C$.  

We simulate the absorbing-state transition model using state-vector evolution, 
as the measurement rate $p_m^A$ increases, the probability of zero fluctuations also increases, indicating a transition from quantum to classical dynamics.

\begin{figure}[htbp]
    \centering
    \includegraphics[width=3.4in]{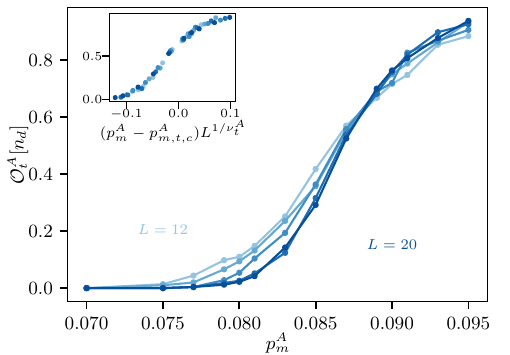}
    \caption{
    Order parameter $\mathcal{O}_{t}^{A}[n_d]$ defined in Eq.~\eqref{eq:op_nd} for trajectory fluctuation using magnetization in Eq.~\eqref{eq:magnetization} 
    for the absorbing-state transition in Ref.~\onlinecite{odea2024entanglement}.
    The system size ranges from $L=12$ to $L=20$ with an evolution time of $t=9L^{z^A}$.
    Inset: Finite-size scaling of the order parameter following Eq.~\eqref{eq:op} with $p_{m,t,c}^A=0.088(2)$  and $\nu_t^A=1.12(2)$. 
    }
    \label{fig:traj_fluct_FSS_APT}
\end{figure}

This universal value implies that the scaling dimension of the weight of zero trajectory fluctuations is zero, which is different from the Bernoulli map model where we find a non-zero scaling dimension $\beta_t=1$.
We thus present the weight of zero fluctuations in the trajectory fluctuations as a function of the measurement rate $p_m^A$ along with the finite-size scaling in Fig.~\ref{fig:traj_fluct_FSS_APT}, following the same definition in Eq.~\eqref{eq:op} as
\begin{equation}\label{eq:op_nd}
    \mathcal{O}_{t}^{A}[n_d] = P[\sigma_{t}^2[n_d]=0]\approx \int_0^\epsilon p[\sigma_{t}^2[n_d]]~ d \sigma_{t}^2[n_d].
\end{equation}
Here, unlike the order parameter in the control transition presented in the main text as shown in Figs.~\ref{fig:traj_fluct_FSS} and~\ref{fig:traj_fluct_FSS_O}, the absorbing-state transition model manifests a universal crossing at the critical point for different system sizes.
Therefore, we can perform the finite-size scaling according to
\begin{equation}
    \mathcal{O}_{t}^{A}[n_d] \sim f_A\left( \left( p_m^A-p_{m,t,c}^A \right)L^{1/\nu_t^A} \right),
\end{equation}
and find a critical measurement rate $p_{m,t,c}^A=0.088(2)$ and a spatial critical exponent $\nu_t^A=1.12(2)$, which is close to the directed percolation in $1.08$ as in Ref.~\cite{odea2024entanglement}. Here, the difference between the Haar random adaptive circuit and the Bernoulli map is that the former has a ``true'' (i.e., exact) absorbing state, so that $\mathcal{O}_{t}^{A}[n_d]$ does not manifest a finite scaling dimension.

\begin{figure}[htbp]
    \centering
    \includegraphics[width=3.4in]{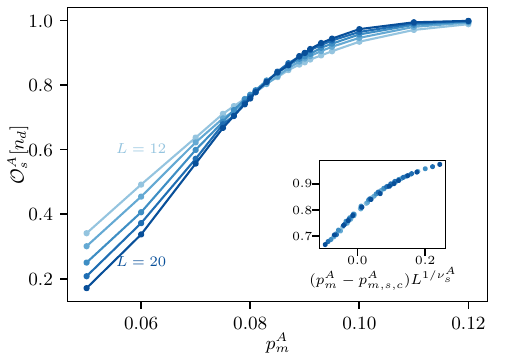}
    \caption{
    Order parameter $\mathcal{O}_{s}^{A}[n_d]$ defined similarly in Eq.~\eqref{eq:op_nd} for trajectory fluctuation using magnetization in Eq.~\eqref{eq:magnetization} 
    for the absorbing-state transition in Ref.~\onlinecite{odea2024entanglement}.
    The full ensemble includes 2000 different circuits $\C$, and 500 different trajectories $\mC$ per circuit.
    The system size ranges from $L=12$ to $L=20$ with an evolution time of $t=3L^{z^A}$.
    Inset: Finite-size scaling of the order parameter following Eq.~\eqref{eq:op} with $p_{m,s,c}^A=0.082(2)$ and $\nu_s^A=1.15(3)$. 
    }
    \label{fig:state_fluct_FSS_APT}
\end{figure}

Similarly, for the state fluctuations in Eq.~\eqref{eq:state_fluct_APT}, we evolve the circuit only up to $t_f = 3 L^{z^A}$ and present the order parameter 
$\mathcal{O}_s[n_d] $ as a function of $p_m^A$ as in Fig.~\ref{fig:state_fluct_FSS_APT}.
We find qualitatively similar behavior, with a critical measurement rate $p_{m,s,c}^A=0.082(2)$ and a spatial critical exponent $\nu_s^A=1.15(3)$.
We also notice a difference in the typical time scale where the trajectory and state fluctuations start to be suppressed: the trajectory fluctuations typically happen at a later time after the state fluctuations are suppressed in the model with exact absorbing states.

\subsection{Bricklayer Clifford random adaptive monitored circuit}
The second generalization is to the monitored adaptive Clifford model studied in Ref.~\cite{sierant2023controlling}, which exhibits an absorbing-state transition that overlaps with the entanglement transition for long range feedback, and separates for short range feedback.
In contrast to the Haar random models, the Clifford model can be efficiently simulated classically, and therefore, we can access larger system sizes.

Here, the absorbing state is still the ferromagnetic state $\ket{0}^{\otimes L}$ in a 1D qubit chain of length $L$.
However, since there is no two-qubit entangling gate in the Clifford group that preserves the state $\ket{00}$, we need to introduce a classical register for each qubit to label whether the qubit is ``active'' (0) or ``inactive'' (1).
Following a bricklayer pattern, at each time step, we first apply a two-qubit projective measurement along the computational basis to each pair of neighboring qubits with the probability $p_m^{\text{cl}}$, and set the classical register of any qubit measured to be `0' to inactive.
After the projective measurements, we adaptively apply a random two-qubit Clifford unitary \textit{only when} at least one of the classical registers of the two qubits is active, after which we set both classical registers to be active.
If both qubits are inactive, we skip the unitary operation such that the ferromagnetic state is a true absorbing state throughout the dynamics.
Finally, we apply a series of variable-range scrambling unitaries among those active qubits with a probability of $1/r^\alpha$ up to a normalization factor, where $r$ is the shortest distance between the qubit pair, and $\alpha$ controls the range of the scrambling.
In the following results, we set $\alpha=0.5$ to study a long-range scenario where the absorbing-state transition is at $p_c = 0.672$ showing a directed procolation universality with the dynamical exponent $z^{\text{cl}}=1.62$, and the correlation length of $\nu^{\text{cl}} = 1.07$. 
We use the same order parameter as in Eq.~\eqref{eq:n_d} to probe the absorbing-state transition.

\subsubsection{Quantum coherence}
We first study the $l_1$-coherence following the definition in Eq.~\eqref{eq:coherence}.
We use the standard tableau representation derived from the Gottesman-Knill theorem~\cite{gottesman1998heisenberg} to efficiently simulate the Clifford dynamics, with the help of the Python package \texttt{stim}~\cite{gidney2021stim}.
The quantum coherence in Eq.~\eqref{eq:coherence} of a stabilizer state can be efficiently obtained from its tableau representation, since only the stabilizers with $X$ and $Y$ operators contribute to the off-diagonal elements in the computational basis, as per
\begin{equation}\label{eq:C_Clifford}
   \mathfrak{C}(\rho) = 2^{\text{rank}(X_\rho) }-1,
\end{equation}
where $X_\rho$ is the $L$ by $L$ matrix for the $X$ sector in the tableau representation of the pure state $\rho$ over $\mathbb{F}_2$. 
For the derivation of Eq.~\eqref{eq:C_Clifford}, we refer the reader to Appendix~\ref{app:C_Clifford}.

\begin{figure}[htbp]
    \centering
    \includegraphics[width=3.4in]{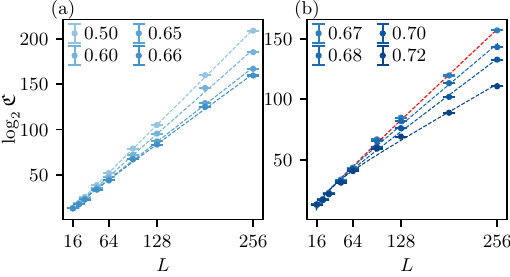}
    \caption{
    Quantum coherence $\mathfrak{C}$ [Eq.~\eqref{eq:coherence} and Eq.~\eqref{eq:C_Clifford}] as a function of $p_m^{\text{cl}}$ in (a) chaotic phase showing exponential growth with system size $L$ using state vector evolution, and (b) the critical (red dashed line) and controlled phase showing sub-exponential growth using matrix product states.
    The evolution time step is set to be $0.02L^{z^{\text{cl}}}$. 
    The dashed line is the fitting of the coherence following Eq.~\eqref{eq:coherence_fiting}. The fitted parameters are shown in Fig.~\ref{fig:coherence_fitting_Clifford}. 
    }
    \label{fig:coherence_L_Clifford}
\end{figure}

We present the quantum coherence of the adaptive Clifford models in Fig.~\ref{fig:coherence_L_Clifford}(a) for the nonabsorbing phase ($p_m^{\text{cl}}<p_{m,c}^{\text{cl}}$) showing the exponential growth with system size $L$, and in Fig.~\ref{fig:coherence_L_Clifford}(b) for the critical and absorbing phase ($p_m^{\text{cl}}\gtrapprox p_{m,c}^{\text{cl}}$) showing sub-exponential growth.

\begin{figure*}[htbp]
    \centering
    \includegraphics[width=6.8in]{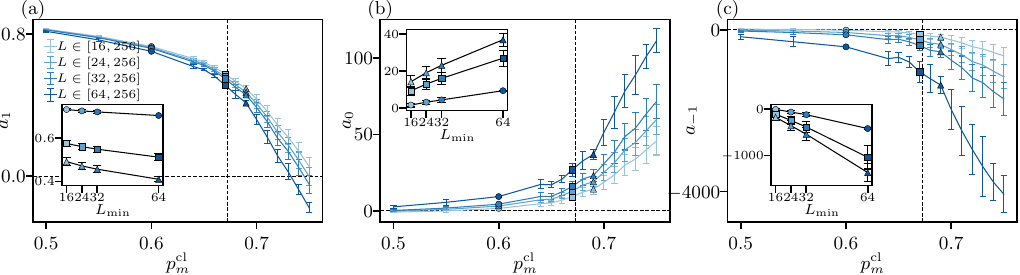}
    \caption{
        The fitted parameters (a) $a_1$, (b) $a_0$, and (c) $a_{-1}$ appearing in the series expansion of the coherence $\mathfrak{C}$ posited in Eq.~\eqref{eq:coherence_fiting}. 
        The different curves correspond to different system size ranges $[L_{\rm min},256]$ with varying $L_{\rm min}$ as shown in the legend.        
        The insets show the trends of the fitted parameters as a function of $L_{\text{min}}$ for $p_m^{\text{cl}}=0.5$ (circles),   $p_m^{\text{cl}}=0.53$ (squares), and $p_m^{\text{cl}}=0.55$ (triangles).
        }
    \label{fig:coherence_fitting_Clifford}
\end{figure*}

Using the same fitting strategy as in Sec.~\ref{sec:coherence}, we fit the log of the average coherence using the series expansion in Eq.~\eqref{eq:coherence_fiting}, and analyze the fitted parameters $a_1$, $a_0$, and $a_{-1}$ in Fig.~\ref{fig:coherence_fitting_Clifford}, where we find similar behaviors as in the Bernoulli map. 


\subsubsection{Quantum fluctuation}
Finally, we present the trajectory and state fluctuations for the adaptive Clifford model using the same definitions in Eqs.~\eqref{eq:traj_fluct_APT} and~\eqref{eq:state_fluct_APT}.
Here, we define the circuit realization $\mathcal C$ by fixing the positions of the measurements, the random local unitaries (if applied), and the realization of the variable-range scrambling gates and the qubits they act on.
The trajectory $\mC$ is associated with a specific sequence of measurement outcomes in the circuit $\C$.

\begin{figure}[htbp]
    \centering
    \includegraphics[width=3.4in]{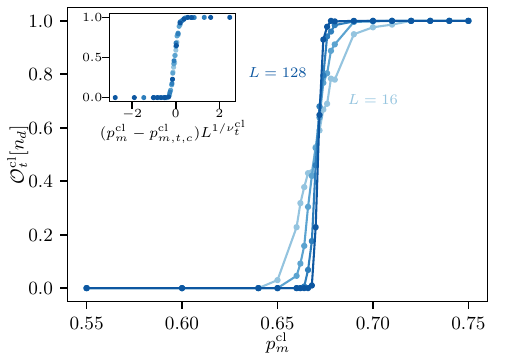}
    \caption{
    Order parameter for trajectory fluctuations $\mathcal{O}_{t}^{\text{cl}}[n_d]$ defined in Eq.~\eqref{eq:op_nd}
    for the adaptive Clifford models in Ref.~\onlinecite{sierant2023controlling}
    The full ensemble includes 500 different circuits $\C$, and 500 different trajectories $\mC$ per circuit.
    The system size ranges from $L=16$ to $L=128$ with an evolution time of $t=6L^{z^{\text{cl}}}$, where $z_{\text{cl}}=1.62$ 
    Inset: Finite-size scaling of the order parameter following Eq.~\eqref{eq:op} with $p_{m,t,c}^{\text{cl}}=0.671(1)$ and $\nu_{t}^{\text{cl}}=1.084(9)$. 
    }
    \label{fig:traj_fluct_FSS_Clifford}
\end{figure}

\begin{figure}[htbp]
    \centering
    \includegraphics[width=3.4in]{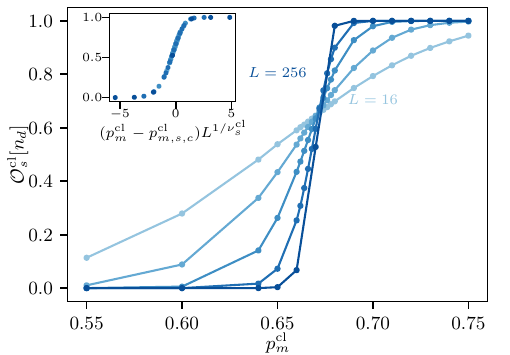}
    \caption{
    Order parameter for state fluctuations $\mathcal{O}_{s}^{\text{cl}}[n_d]$ defined in Eq.~\eqref{eq:op_nd} for the adaptive Clifford models in Ref.~\onlinecite{odea2024entanglement}
    The full ensemble includes 500 different circuits $\C$, and 500 different trajectories $\mC$ per circuit.
    The system size ranges from $L=16$ to $L=128$ with an evolution time of $t=L^{z^{\text{cl}}}$, where $z_{\text{cl}}=1.62$ 
    Inset: Finite-size scaling of the order parameter following Eq.~\eqref{eq:op} with $p_{m,s,c}^{\text{cl}}=0.671(1)$ and $\nu_{s}^{\text{cl}}=1.08(6)$. 
    }
    \label{fig:state_fluct_FSS_Clifford}
\end{figure}

We present the order parameters for the trajectory fluctuations $\mathcal{O}_{t}^{\text{cl}}[n_d]$ in Fig.~\ref{fig:traj_fluct_FSS_Clifford} and for the state fluctuations $\mathcal{O}_{s}^{\text{cl}}[n_d]$ in Fig.~\ref{fig:state_fluct_FSS_Clifford}, following the definition in Eq.~\eqref{eq:op_nd}. 
We find both order parameters show a universal crossing at the critical point, and the finite size scaling following Eq.~\eqref{eq:fss} gives similar critical exponents with $\nu_{t}^{\text{cl}}=1.084(9)$ for the trajectory fluctuations and $\nu_{s}^{\text{cl}}=1.08(6)$ for the state fluctuations.

\section{Discussion}\label{sec:discussion}
In this work, we have provided a framework to characterize the quantum nature of CIPTs in the presence of unitaries, measurements, and feedback. 
The classical contributions have been precisely identified and, in the case of the Bernoulli circuit model, have been shown to nicely capture the average long-time steady state properties, including the universal properties of the underlying classical absorbing state transition. 
By separating out the classical and quantum contributions to an average over measurement outcomes as in Eq.~\eqref{eq:decompose}, we have been able to identify a critical mode associated with trajectory-to-trajectory fluctuations governed by a universal scaling dimension unique to the quantum limit.
These critical fluctuations among the many ``worlds" of quantum mechanics can be directly measured on NISQ hardware without the need for post-selection. 
To distinguish quantum from classical fluctuations in such experiments, it will be crucial to extend the ideas discussed in this paper to include the effects of noise present on NISQ devices.

\HPadd{We also analyzed the quantum coherence across the CIPT using a combination of exact state-vector evolution, matrix product state simulations leveraging tensor cross interpolation, and Clifford circuit simulations.
}By uncovering the critical behavior of the coherence at the CIPT, we have shown that absorbing-state transitions embedded into quantum systems also entail a phase transition between quantum coherent and fully classical dynamics.

\HPadd{Our method to unveil the quantum nature of the absorbing-state transition induced by measurement with feedback, implied by both the exponential growth of quantum coherence at the critical point and the critical behavior of the weight of zero trajectory/state fluctuations of any observable, can be also generalized to other types of absorbing-state transitions beyond the Bernoulli map.
We have demonstrated this explicitly for two additional models: the Haar random adaptive circuit~\cite{odea2024entanglement} and a Clifford random adaptive circuit~\cite{sierant2023controlling}.
Interestingly, both models exhibit a critical behavior in the trajectory and state fluctuations showing a universal crossing, and do not manifest a finite scaling dimension as in the Bernoulli map. We conjecture that this difference is due to the nature of the absorbing state: while the Bernoulli map has an approximate absorbing state, the Haar random and Clifford models have an exact absorbing state throughout the dynamics. 

}

In all models we have investigated in this work, the phase transition in trajectory-to-trajectory fluctuations and quantum coherence coincide. 
However, this need not be the case in general.
For example, if we were to instead consider controlling onto an area-law entangled state, such as the Greenberger-Horne-Zeilinger or cluster states as in Ref.~\cite{odea2024entanglement}, the two transitions may split apart. 
Understanding the structure of these critical properties in a precise manner is now possible utilizing the model-independent framework developed in this work.
Opening this up to target states with various patterns of entanglement and correlations is a fascinating direction for future research.

In future work, it will be essential to incorporate the effects of noise into this formalism for several reasons. \HPadd{
In Ref.~\cite{pokharel2025order}
we have verified using a classical dephasing model that certain quantum features (such as the full fluctuation structure) cannot be ``spoofed'' by classical mixtures that retain circuit-level randomness while removing phase coherence. However, a more comprehensive study is needed to understand} whether the CIPT is stable to realistic noise channels, or if it rounds the transition into a crossover similar to the MIPT~\cite{li2023statistical}. 
Moreover, to understand how the critical quantum fluctuations discussed in this work can be observed experimentally, it is essential to characterize the effects of noise on each fluctuation parameter we have introduced. For example, focusing on the tail of the distribution of fluctuations close to zero is likely ill suited for experimental application, as the smallest fluctuations will be strongly affected by noise and may not allow one to discern quantum from classical fluctuations. \jp{Therefore, in the recent observation of the Bernouli CIPT on IBM's superconducting processor (by three of the authors), we utilized moments of the magnetization, which allowed us to seperate quantum and classical fluctuations provided $p_{\mathrm{ctrl}}<p_c$. Extensions to observables, and a more comprehensive analysis of the noise we leave these generalizations for future work.}

\section*{Acknowledgments}
We thank David A. Huse and Justin Wilson for their valuable feedback on versions of this manuscript, as well as Kemal Aziz, Abhinav Deshpande, Sriram Ganeshan, Luke Govia, Barbara Jones, Bibek Pokharel, Oles Shtanko, Maika Takita, and Justin Wilson for many valuable discussions and for collaboration on related work.
This work is partially supported by US-ONR grant No.~N00014-23-1-2357 (H.P. and J.H.P.) and by the National Science Foundation under Grants No.~DMR-2143635 and No.~DMR-2611305 (T.I.).
\bibliography{Paper_CIPT.bib}

\appendix

\section{Distribution of the bit string and the first domain wall}\label{app:distribution}
\begin{figure}[htbp]
    \centering
    \includegraphics[width=3.4in]{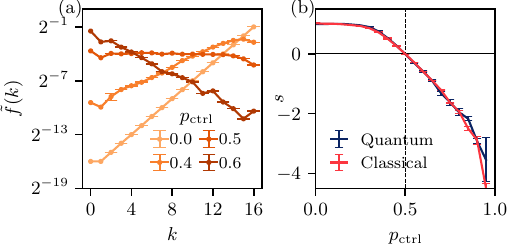}
    \caption{
    (a) Ensemble-averaged distribution of the FDW location $k$ with $L=16$ for the quantum model.
    (b) The fitted exponent $s$ characterizing the FDW distribution as a function of $p_{\text{ctrl}}$ for the quantum (blue) and the classical (red) models.
    The error bars are estimated from an ensemble size of 2000 for each $p_{\text{ctrl}}$.
        }
    \label{fig:FDW}
\end{figure}

In Sec.~\ref{sec:bitstring}, we introduce the distribution of the bit string $x$.
In this section, we show that it is directly related to the distribution of the first domain wall $k$, which is exponential distribution with an exponent $s$ which can be numerically extracted in Fig.~\ref{fig:FDW}(b).

With the concept of FDW, it is convenient to rewrite the distribution of bit string in to a more compact form by grouping bit strings according to their FDW location $k$ and plot the cumulative distribution $\tilde f(k)$. In Fig.~\ref{fig:FDW}(a), we find that $\tilde f(k)$ takes an exponential form for both the classical and quantum models. 
To analyze the behavior of $\tilde f(k)$ as a function of $p_{\rm ctrl}$, we define the ansatz
\begin{equation}\label{eq:FDW_ansatz}
    \tilde{f}(k)\equiv \sum_{x\in \left\{ x|\text{FDW}(x)=k \right\}}f(x) \approx 2^{k} f(x) \propto  2^{sk},
\end{equation}
where the exponent $s$ is a fitting parameter that characterizes the distribution.
When $s<0$ ($s>0$), the FDW is localized at $k=0$ ($k=L$), reflecting that the system is in the controlled (chaotic) phase.
As $p_{\text{ctrl}}$ increases from zero, the exponent $s$ passes through zero at the CIPT, as shown in Fig.~\ref{fig:FDW}(b). 

With the exponent $s$ known, we can further analytically derive the scaling behavior of the distribution of the bit string $x$ and the first domain wall $k$.

At zero control rate $p_{\text{ctrl}}=0$, the distribution of the FDW $\tilde{f}(k)$ is simply proportional to the number of bit strings within same FDW $k$, i.e., $2^{k}$. Therefore, $s$ should be 1 as also numerically verified in Fig.~\ref{fig:FDW}(a).

As $p_{\text{ctrl}}$ increases, the linearity remains but with a smaller slope below 1 until $p_{\text{ctrl}}$ approaches the critical rate $p_{\text{ctrl}}=0.5$, which gives a zero slope $s=0$, indicating a uniform distribution as a function of the FDW $k$, i.e., $\tilde{f}(k)\propto 1/L$.
The uniform distribution of $\tilde{f}(k)$ at $p_{\text{ctrl}}=0.5$ indicates that the bit strings $x$ start to localize at the fixed point $x_f=0$ as the distribution of the bit strings $f(x)\propto 2^{-k}/L$.

Beyond the critical point $p_{\text{ctrl}}=0.5$, the distribution of FDW is linear with a negative slope, implying a distribution of the FDW as $\tilde{f}(k)\propto 2^{sk}$ with $s<0$.
This indicates an exponential localization at the fixed points with $k=0$ for $p_{\text{ctrl}}>0.5$ compared to the algebraic decay at $p_{\text{ctrl}}=0.5$. At the limit of control $p_{\text{ctrl}}=1$, the slope $s\rightarrow-\infty$ indicates that the fixed points are the only possible states.

In summary, considering the normalization factor, the complete description of the distribution of the bit string $x$ in Eq.~\eqref{eq:fidelity} 
\begin{equation}\label{eq:fx}
    f(x)=2^{-k} \tilde{f}(k) 
\end{equation}
where
\begin{equation}\label{eq:FDW}
    \tilde{f}(k) =\begin{cases}
        \frac{1-2^s}{1-2^{s(L+1)}}2^{sk} & s\neq 0 ~(p_{\text{ctrl}}\neq 0.5)\\
        \frac{1}{L+1} & s=0 ~ (p_{\text{ctrl}}=0.5)\\
    \end{cases},
\end{equation}
for $k\in \left[ 0,L \right]$ being the position of FDW for the bit string $x$.
Here, the only unknown parameter is the slope $s$ where it changes from $+1$ at $p_{\text{ctrl}}=0$ to 0 at $p_{\text{ctrl}}=0.5$ and to $-\infty$ at $p_{\text{ctrl}}=1$, which can be fitted numerically in Fig.~\ref{fig:FDW}(b). 
We also note that the slope $s$ converges for both the classical and quantum models, which reiterates the fact that the quantum model inherits the classical critical point.

\begin{figure}[htbp]
    \centering
    \includegraphics[width=3.4in]{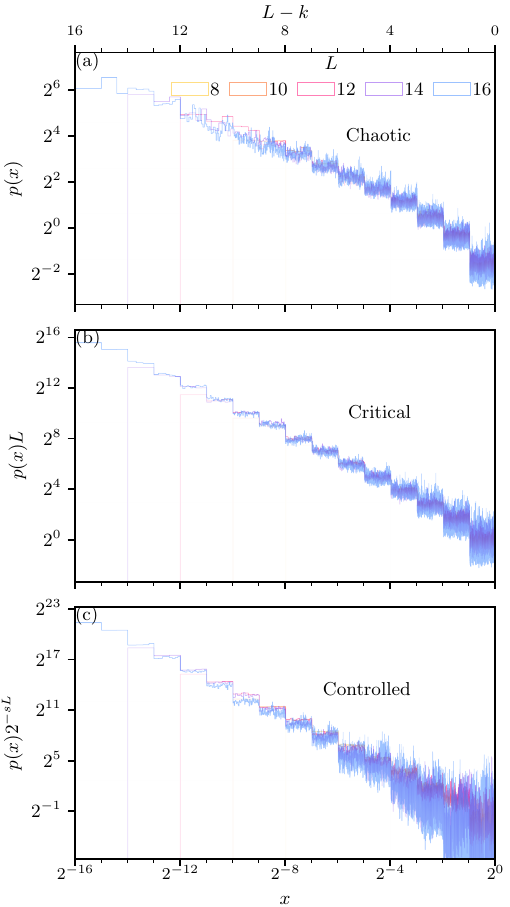}
    \caption{
    The probability density function of the bit string for (a) the chaotic phase at $p_{\text{ctrl}}=0.4$, (b) the critical point at $p_{\text{ctrl}}=0.5$, and (c) for the controlled phase at $p_{\text{ctrl}}=0.6$. 
    Note that the different scale in the vertical axis, which corresponds to the scaling in (a) Eq.~\eqref{eq:px_chaotic}, (b) Eq.~\eqref{eq:px_critical}, and (c) Eq.~\eqref{eq:px_control}, such that they all collapse to the same curve.
    }
    \label{fig:tail_distribution}
\end{figure}
\subsection{Scaling of the tail}
With the analytic description of the distribution of the bit string, we can fully understand the drastic change in controllability before and after the critical point by studying the distribution of the tail (i.e., $\left\{ x|x\neq x_f \right\}$) in Fig.~\ref{fig:tail_distribution}.
We show the distribution of the tail for different system sizes $L$ from the chaotic phase ($p_{\text{ctrl}}=0.4$ in Fig.~\ref{fig:tail_distribution}(a)), to criticality ($p_{\text{ctrl}}=0.5$ in Fig.~\ref{fig:tail_distribution}(b)), and finally to controlled phase ($p_{\text{ctrl}}=0.6$ in Fig.~\ref{fig:tail_distribution}(c)).

In order to compare different system sizes on an equal footing, we define the probability density function $p(x)$ over a continuous variable $x$, which can be considered as an extrapolation to the thermodynamic limit.
The probability density function $p(x)$ is obtained by factoring the measure of a single bit string (i.e., $2^{-L}$). 
Namely, 
\begin{equation}\label{eq:px}
    f(x)= p(x)dx\approx p(x)\Delta x= 2^{-L}p(x).
\end{equation}

In the chaotic phase as shown in Fig.~\ref{fig:tail_distribution}(a), we find that the probability density function $p(x)$ converges to an $L$-independent distribution,
which can be also understood from the analytic result Eq.~\eqref{eq:px} and Eq.~\eqref{eq:FDW} with  
\begin{equation}\label{eq:px_chaotic}
    \begin{split}
        p(x) &= f(x) 2^{L}=  2^{-k} \frac{1-2^s}{1-2^{s(L+1)}}2^{sk} 2^{L} \\
        &\approx x^{s-1}(2^{s}-1)
    \end{split}
\end{equation}
Here, the last line is obtained because $s\in\left( 0,1 \right]$ for $p_{\text{ctrl}}\in\left[ 0,0.5 \right)$ in the thermodynamic limit, and by noting Eq.~\eqref{eq:k}.

At the critical point $p_{\text{ctrl}}=0.5$ as shown in Fig.~\ref{fig:tail_distribution}(b), the probability density function $p(x)$ decays algebraically as $L$ increases because, 
\begin{equation}\label{eq:px_critical}
    \begin{split}
        p(x) &= f(x) 2^{L}=2^{-k}\frac{1}{L} 2^L\\
        &\approx \frac{1}{2(L+1)x},
    \end{split}
\end{equation} 
This can also be visually confirmed in Fig.~\ref{fig:tail_distribution}(b) that $p(x)L$ is an exponential distribution independent of $L$.

Finally, in the controlled phase as shown in Fig.~\ref{fig:tail_distribution}(c), the probability density function $p(x)$ decays exponentially as $L$ increases for the tail while diverging at the fixed points in the thermodynamic limit because
\begin{equation}\label{eq:px_control}
    \begin{split}
        p(x)&=f(x)2^{L}=2^{-k}\frac{1-2^s}{1-2^{s(L+1)}}2^{sk}2^{L}\\
            &\approx x^{s-1} 2^{Ls}        
    \end{split}
\end{equation}
where $s\in\left( -\infty, 0 \right)$ for $p\in\left( 0.5,1 \right]$.
Therefore, in the thermodynamic limit, any non-fixed point bit string will be exponentially suppressed in the controlled phase, which indicates a Dirac $\delta$-distribution for the fixed point, i.e., $\delta(x)$.
This means that all the dynamics will be steered to the fixed point given a sufficient rate of control maps ($p_{\text{ctrl}}>0.5$) in the thermodynamic limit.
The $L$-independent exponential decay of $p(x)2^{-sL}$ in Fig.~\ref{fig:tail_distribution}(c) also confirms the analytic result Eq.~\eqref{eq:px_control}.

\section{Quantum coherence for a Haar random state}\label{app:coherence_Haar}
In this section, we analytically derive the $l_1$-coherence of a Haar random state, denoted as
\begin{equation}\label{eq:psi}
    \ket{\psi}=\sum_{x=1}^{N} z_x \ket{x}=\sum_{x=1}^{N} (a_x+i b_x) \ket{x},
\end{equation}
where $N=2^L$ is the total number of basis, and $a_x,b_x\in\mathbb{R}$ are uniformly distributed on the surface of a unit sphere, i.e., $\sum_x a_x^2+b_x^2=1$.

Following Eq.~\eqref{eq:coherence}, the $l_1$-coherence for $\ket{\psi}$ averaged over the Haar random ensemble is
\begin{equation}\label{eq:coherence_Haar}
    \begin{split}
        \mathfrak{C}&=\expval{\sum_{x\neq x'}\abs{\rho_{x,x'}}}_{\text{Haar}}\\
        &=\sum_{x\neq x'}\expval{\sqrt{a_x^2+b_x^2}\sqrt{a_{x'}^2+b_{x'}^2}}_{\text{Haar}}\\
        &=N(N-1)\expval{\sqrt{a_1^2+b_1^2}\sqrt{a_{2}^2+b_{2}^2}}_{\text{Haar}}\\
        &=N(N-1)\int_{\sum_x a_x^2+b_x^2=1} \prod_{x=1}^{N} da_x db_x \sqrt{a_1^2+b_1^2}\sqrt{a_{2}^2+b_{2}^2}\\
        &=N(N-1)\int \prod_{x=1}^{2N-1} \mathcal{D}\theta_x \sqrt{\cos^2\theta_1+\sin^2\theta_1\cos^2\theta_2}\\
        &\sin\theta_1\sin\theta_2\sqrt{\cos^2\theta_3+\sin^2\theta_3\cos^2\theta_4},  
    \end{split}
\end{equation}
where the new measure in $\theta_x$ is 
\begin{equation}
    \mathcal{D}\theta_x=\prod_{x=1}^{2N-1}\sin^{2N-x-1}\theta_x~d\theta_x,
\end{equation}
from the polar coordinate on a hypersurface defined as
\begin{equation}\label{eq:jacobian}
    a_x = \prod_{j=1}^{2x-2}\sin\theta_j \cos\theta_{2x-1},~b_x =\prod_{j=1}^{2x-1}\sin\theta_j\cos\theta_{2x}.
\end{equation} 
Here, each $\theta_x$ integrates from $0$ to $\pi$ except for the last one $\theta_{2N-1}$ integrates from $0$ to $2\pi$. (Note that $b_{N}$ is not independent, so we only have $\theta_1$ to $\theta_{2N-1}$.)
The third equal sign is due to the cyclic symmetry under the permutation of $a_x$ and $b_x$.

The last line in Eq.~\eqref{eq:coherence_Haar} can be decomposed into the product of the following three integrals 
\begin{equation}
        \int \sqrt{\cos^2\theta_1+\sin^2\theta_1\cos^2\theta_2}\prod_{x=1}^2\sin\theta_x\mathcal{D}\theta_x=\frac{(2N-3)!!}{(2N)!!}\pi^2,
\end{equation}
\begin{equation}
    \int \sqrt{\cos^2\theta_3+\sin^2\theta_3\cos^2\theta_4}\prod_{x=3}^4\mathcal{D}\theta_x =\frac{(2N-6)!!}{(2N-3)!!}2\pi,
\end{equation}
and the normalization factor
\begin{equation}
    \int \prod_{x=5}^{2N-1}\mathcal{D}\theta_x=\left( \int \prod_{x=1}^{4}\mathcal{D}\theta_x \right)^{-1}= \frac{(2N-2)(2N-4)}{4\pi^2}.
\end{equation}
This leads to each off-diagonal element contributing on average
\begin{equation}
    \expval{\sqrt{a_1^2+b_1^2}\sqrt{a_{2}^2+b_{2}^2}}_{\text{Haar}}=\frac{\pi}{4N},
\end{equation}
and therefore, the $l_1$-coherence averaged over the total number of basis is
\begin{equation}
    \frac{1}{N}\mathfrak{C} = \frac{1}{N}\frac{\pi}{4N}N(N-1)=\frac{\pi(N-1)}{4N}.
\end{equation}
Since $N$ is exponentially large to system size $L$, in the thermodynamic limit, the $l_1$-coherence of a Haar random state per basis is $\frac{\pi}{4}$.

\section{Tensor cross interpolation}\label{app:TCI}
In this section, we provide parameters on the tensor cross interpolation method used in the main text to compute the $l_1$-coherence from the MPS.
Since we always stay in the area-law phase, we can ensure that the bond dimension of the MPS will not grow exponentially with the system size $L$. 
Therefore, for the circuit evolution, we only set the truncation error as $10^{-10}$ without restricting the allowed maximal bond dimension.

Although the efficient simulation of the circuit using MPS is possible, the tensor cross interpolation of the final wave function to compute the $l_1$-coherence, which requires $O(\chi^4)$ resources, is not guaranteed for large system sizes \textit{a priori}.
Therefore, we first validate the tensor cross interpolation method using the package `TensorCrossInterpolation.jl'~\cite{2025tensor4all} to estimate the $l_1$-coherence in small systems sizes ($L\le20$) by comparing the results with the exact state-vector evolution to find consistency. 

For larger system sizes, we rely on the following error analysis. 
Since the TCI essentially is an interpolation method, meaning that for each bit string $x=b_1b_2\dots b_L$, we can evaluate its deviation from the exact value (e.g., the norm of the wave function amplitude) by contracting all internal indices.
We randomly sample a few thousands different bit strings, and take the largest difference between the exact and interpolated values, which provides an estimate of upper bound of the error of the tensor cross interpolation, denoted as $\delta(\chi,L)$ given a maximal bond dimension $\chi$ in TCI and a system size $L$.

Finally, we can estimate the propagation of error from TCI to the $l_1$-coherence of a specific pure state $\rho$, given by
\begin{equation}\label{eq:C_simp}
    \mathfrak{C}(\rho)= \sum_{x\neq x'}\abs{\rho_{x,x'}}=\left( \sum_x \abs{z_x} \right)^2-1,
\end{equation}
where $z_x$ is the wave function amplitude of the pure state $\rho=\ketbra{\psi}$ at a given basis $x$ (see Eq.~\ref{eq:psi}).
{Here, to efficiently compute the summation in the last equation in Eq.~\eqref{eq:C_simp} (i.e., $ \sum_x \abs{z_x} $), we take the inner product of the interpolated MPS representing the vector $\left( \abs{z_0},\abs{z_1},\dots,\abs{z_{2^L-1}} \right)^\intercal$ with the all-one vector of the same dimension.}
With the upper bound on error $\delta(\chi,L)$ in $\abs{z_x}$, we can estimate the final error of the $l_1$-coherence by substituting $\abs{z_x}\rightarrow {\abs{z_x}+\delta(\chi,L)}$, which gives the maximal error of the $l_1$-coherence as $2\sum_x\abs{z_x}\delta(\chi,L)$.
{Therefore, we control the error of the $l_1$-coherence to be $O(1)$ (i.e., the relative error is bounded by $O(e^{-L})$) by increasing the allowed maximal bond dimensions in the TCI, and finally, find empirically the allowed maximal bond dimensions are $\chi_{\text{max}}=30,60,400,1200$, and 4000 for $L=10,20,30,40$, and 50, respectively.}

\section{Fluctuations of the magnetization density}\label{app:fluctuation_O}
\begin{figure}[htbp]
    \centering
    \includegraphics[width=3.4in]{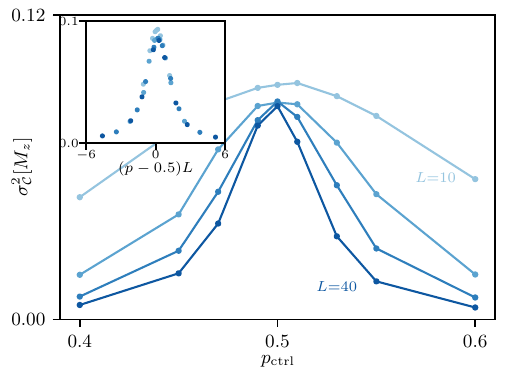}
    \caption{
    Steady-state circuit-to-circuit fluctuations of the magnetization in Eqs.~\eqref{eq:cl_fluct_O} and~\eqref{eq:magnetization} at $p_{\text{ctrl}}=0.5$ for different system sizes $L$.
    The inset shows finite-size-scaling collapse following Eq.~\eqref{eq:fss_sigma_c} with $\nu=0.94(4) $  and $p_c^{\text{CIPT}}=0.50(2)$.
    }
    \label{fig:cl_fluct_FSS_O}
\end{figure}
In this section, we additionally present the fluctuations of the magnetization density $M_z$ defined as
\begin{equation}
    M_z=\frac{1}{L}\sum^L_{i=1}Z_i,
\end{equation}
serving as a `generic' order parameter, to confirm the universality of the scaling dimension of these quantum fluctuations
by demonstrating the same critical exponents presented in the main text.

\subsection{Circuit-to-circuit fluctuations}
The circuit-to-circuit fluctuations of the magnetization density are defined similarly to Eq.~\eqref{eq:cl_fluct} as
\begin{equation}\label{eq:cl_fluct_O}
    \sigma_{\C}^2[M_z]
    =\mathbb{E}_{\C} \left[ \left(\mathbb{E}_{\mC}\left[ ({M_z})_{\mC} \right] \right)^2 \right] - \left( \mathbb{E}_{\C} \mathbb{E}_{\mC} \left[ ({M_z})_{\mC} \right] \right)^2.
\end{equation}
We present the results for the circuit-to-circuit fluctuations in Fig.~\ref{fig:cl_fluct_FSS_O}, and find the same critical exponent as that in Fig.~\ref{fig:cl_fluct_FSS}.

\subsection{Fluctuations across quantum trajectories}
\begin{figure*}[htbp]
    \centering
    \includegraphics[width=6.8in]{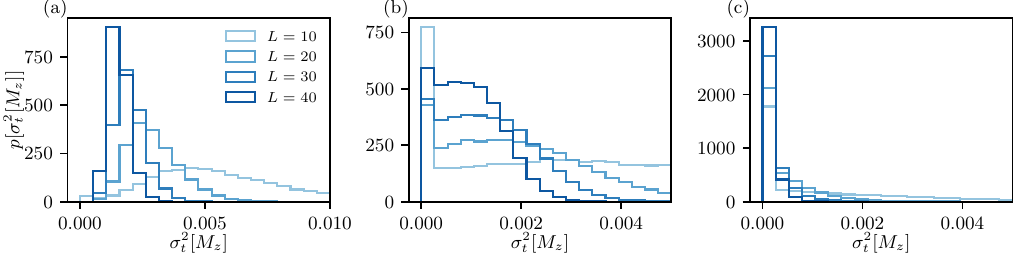}
    \caption{
    Probability distribution over all circuits $\C$ of the trajectory fluctuations $\sigma_{t}^2[M_z]$ in the magnetization $M_z$ in Eq.~\eqref{eq:traj_fluct_O} for steady states at (a) $p_{\text{ctrl}}=0.4$, (b) $p_{\text{ctrl}}=0.5$, and (c) $p_{\text{ctrl}}=0.6$.
    }
    \label{fig:traj_fluct_O}
\end{figure*}
\begin{figure}[htbp]
    \centering
    \includegraphics[width=3.4in]{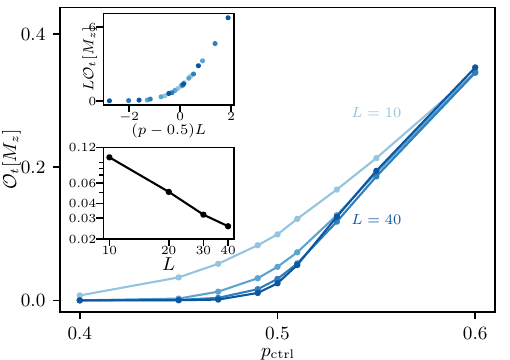}
    \caption{
    Order parameter $\mathcal{O}_{t}[M_z]$ for the trajectory fluctuation of magnetization density $M_z$ [Eq.~\eqref{eq:op} $\sigma_{t}^2[M_z]$] for different system sizes $L$.
    The full ensemble includes 500 different circuits $\C$, and 500 different trajectories $\mC$ per circuit.
    Upper inset: Finite-size scaling of the order parameter following Eq.~\eqref{eq:op} with $\nu=0.91(4)$, $p_c^{\text{CIPT}}=0.498(1)$, and $\beta_t=0.98(4)$.
    Lower inset: Algebraic decay of the order parameter as a function of $L$ at the critical point.
    }
    \label{fig:traj_fluct_FSS_O}
\end{figure}
The fluctuations across quantum trajectories of the magnetization density are defined similarly to Eq.~\eqref{eq:traj_fluct} as
\begin{equation}\label{eq:traj_fluct_O}
    \sigma_{t}^2[M_z] = \mathbb{E}_{\mC} \left[ \expval{M_z}_{\mC}^2 \right] - \left( \mathbb{E}_{\mC} \left[ ({M_z})_{\mC} \right] \right)^2.
\end{equation}
Here, we also find the same transition in the distribution of the fluctuations across quantum trajectories from the chaotic to the controlled phase, as shown in Fig.~\ref{fig:traj_fluct_O}, where the probability of zero fluctuations increases, and plot the probability of zero fluctuations as shown in Fig.~\ref{fig:traj_fluct_FSS_O}.
We find a similar {scaling dimension} as in Fig.~\ref{fig:traj_fluct_FSS}.

\subsection{Fluctuations across quantum states}
\begin{figure}[htbp]
    \centering
    \includegraphics[width=3.4in]{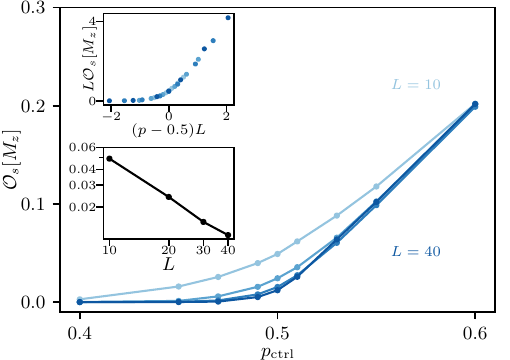}
    \caption{
    Order parameter $\mathcal{O}_s[M_z]$ for the state fluctuation of magnetization density $M_z$ (defined using $\sigma_{s,\mC}^2[M_z]$) for different system sizes $L$ as a function of $p_{\rm ctrl}$. 
    Upper inset: Finite-size scaling collapse with $\nu=0.99(4)$, $p_c^{\text{CIPT}}=0.500(2)$, and $\beta_{s}=1.00(7)$.
    Lower inset: Algebraic decay of the order parameter as a function of $L$ at the critical point.
    }
    \label{fig:state_fluct_FSS_O}
\end{figure}
We also verify the fluctuations across quantum states 
\begin{equation}\label{eq:state_fluct_O}
    \sigma_{s,\mC}^2[M_z]=  \expval{M_z^2}_{\mC} - \expval{M_z}_{\mC}^2 
\end{equation}
and present similar results of finite-size scaling of the probability of zero fluctuations as shown in Fig.~\ref{fig:state_fluct_FSS_O}. 

\subsection{Fluctuations across different shots}\label{app:shots}
\begin{figure}
    \centering
    \includegraphics[width=3.4in]{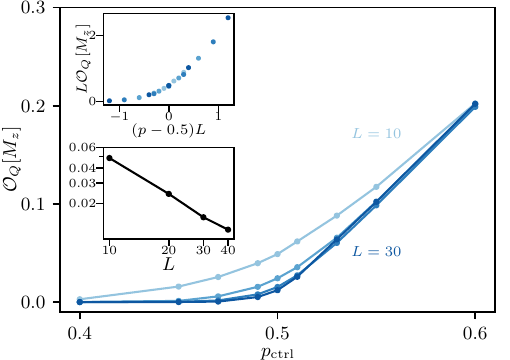}
    \caption{
    Order parameter $\mathcal{O}_{Q}[M_z]$ for the quantum fluctuation of magnetization density $M_z$ [Eq.~\eqref{eq:op} with $\sigma_{Q,\C}^2[M_z]$] for different system sizes $L$.
    The full ensemble includes 500 different circuits $\C$, and 250000 different shots per circuit.
    Upper inset: Finite-size scaling of the order parameter following Eq.~\eqref{eq:op} with $\nu=1.00(7)$, $p_c^{\text{CIPT}}=0.500(5)$, and $\beta_{Q}=1.000(1)$.
    Lower inset: Algebraic decay of the order parameter as a function of $L$ at the critical point.
        }
    \label{fig:shots_fluct_FSS_O}
\end{figure}

Finally, we study the total quantum fluctuation of the magnetization density. 
We define the order parameter $\mathcal{O}_{Q}$, the probability of zero quantum fluctuations, similarly, and find that it exhibits the same critical behavior and exponents as those of the trajectory and state fluctuation, as shown in Fig.~\ref{fig:shots_fluct_FSS_O}.

\section{Finite-size scaling of magnetization density using MPS}\label{app:FSS}
\begin{figure}
    \centering
    \includegraphics[width=3.4in]{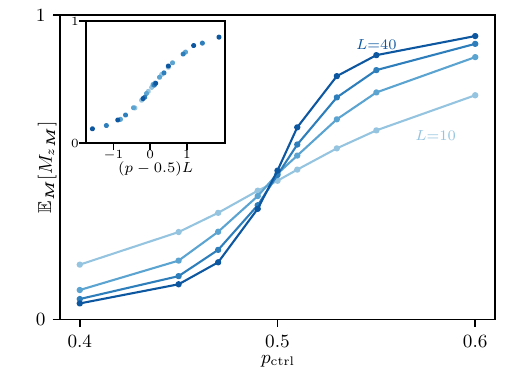}
    \caption{The steady-state ensemble-averaged magnetization density $\mathbb{E}_{\bm{M}}[\expval{M_z}_{\bm{M}}]$ near the critical point of control transition. Inset shows the finite-size scaling following Eq.~\ref{eq:FSS_Mz} with $\nu=1.04(2)$ and $p_c^{\text{CIPT}}=0.496(4)$.}
    \label{fig:O_mean_FSS}
\end{figure}
In this section, we present a better estimate of the critical exponents at the CIPT with larger system sizes up to $L=40$. 
We focus on the steady-state ensemble-averaged magnetization density $\mathbb{E}_{\bm{M}}[({M_z})_{\bm{M}}]$, and present the results in Fig.~\ref{fig:O_mean_FSS}. 
The critical exponents at the transition $p_{\text{ctrl}}=0.5$ can be better estimated under larger system sizes following 
\begin{equation}\label{eq:FSS_Mz}
    \mathbb{E}_{\bm{M}}[({M_z})_{\bm{M}}] \sim f_{M_z}\left( \left( p_{\text{ctrl}}-p_c^{\text{CIPT}} \right)L^{1/\nu} \right),
\end{equation}
where $f_{M_z}$ is a universal scaling function for the ensemble-averaged total magnetization density.
We present the data collapse in the inset of Fig.~\ref{fig:O_mean_FSS}, and extracted a critical exponent of $\nu=1.04(2)$ and $p_c^{\text{CIPT}}=0.496(4)$.

\section{$l_1$-coherence for stabilizer states}\label{app:C_Clifford}
The coherence for a stabilizer state can be efficiently computed within the tableau representation. Following the definition in Eq.~\eqref{eq:C_simp}, the $l_1$-coherence for a pure stabilizer state $\rho$ only depends on the amplitude magnitude of the wave function in the computational basis 
\begin{equation}\label{eq:z_x}
    \abs{z_x} = \sqrt{\rho_{x,x}}.
\end{equation}

For a pure stabilizer state of $L$ qubits, the density matrix in the computational basis can be expressed as 
\begin{equation}
    \rho = \frac{1}{\abs{S}}\sum_{s\in S} s,
\end{equation}
where $S$ is the stabilizer group, and $\abs{S}=2^L$ is the number of elements in $S$. 
We define the subgroup $S_Z = S \cup \left\{ I,Z \right\}^{\otimes L}$ whose elements only contain $I$ and $Z$ operators in the stabilizer Pauli string, and the density matrix can be decomposed into 
\begin{equation}\label{eq:rho_S}
    \rho = \frac{1}{\abs{S}}\left( \sum_{s\in S_Z} s + \sum_{s\notin S_Z} s \right),
\end{equation}
To compute the coherence in Eqs.~\eqref{eq:C_simp} and~\eqref{eq:z_x}, we need the diagonal elements
\begin{equation}\label{eq:rho_xx}
    \rho_{x,x} = \frac{1}{\abs{S}}\sum_{s\in S_Z} \expval{s}{x} ,
\end{equation} 
where the second term in Eq.~\eqref{eq:rho_S} vanishes because any Pauli string containing $X$ or $Y$ operators will flip at least one qubit in the computational basis.

Therefore, from Eq.~\eqref{eq:C_simp}, the $l_1$-coherence for a stabilizer state can be simplified to
\begin{equation}\label{eq:C_rho_xx}
    \mathfrak{C}(\rho) = \left( \sum_{x \in C_{S_Z}} \sqrt{\rho_{x,x}} \right)^2-1,
\end{equation}
where $C_{S_Z} = \left\{ x: s \ket{x} = \ket{x}, \forall s \in S_Z \right\}$ is the computational basis stabilized by the subgroup $S_Z$, leading to 
\begin{equation}\label{eq:rho_xx_val}
    \rho_{x,x} = \frac{\abs{S_Z}}{\abs{S}},
\end{equation} 
from Eq.~\eqref{eq:rho_xx}. 
Other computational basis outside $C_{S_Z}$ will be manifestly zero due to the group symmetry. 
(Because if $x\notin C_{S_Z}$, it means there exists at least one $s'\in S_Z$ such that $s'\ket{x}=-\ket{x}$. Therefore, $ -\rho_{x,x}=\expval{s'}{x}\rho_{x,x} = \sum_{s\in S_Z} \expval{s'}{x}\expval{s}{x} =  \sum_{s\in S_Z} \expval{s's}{x}= \rho_{x,x}$ leads to $\rho_{x,x}=0$, where the third equal sign is because $s'$ and $s$ are both diagonal in computational basis)

Due to the normalization of the density matrix, we have 
\begin{equation}
    \sum_{x\in C_{S_Z}} \rho_{x,x} = \abs{C_{S_Z}} \frac{\abs{S_Z}}{\abs{S}} = 1,
\end{equation}
which gives the size of the set (also the number of nonzero wave function amplitudes)
\begin{equation}\label{eq:C_S_Z}
    \abs{C_{S_Z}} = \frac{\abs{S}}{\abs{S_Z}}.
\end{equation}

Finally, substituting Eq.~\eqref{eq:rho_xx_val} and Eq.~\eqref{eq:C_S_Z} back to the $l_1$-coherence in Eq.~\eqref{eq:C_rho_xx}, we obtain
\begin{equation}
    \mathfrak{C}(\rho) =  \frac{\abs{S}}{\abs{S_Z}}-1=2^{L-r_Z}-1,
\end{equation}
where $r_Z$ is the number of generators of the subgroup $S_Z$.

To count the size of $S_Z$ ($\abs{S_Z}=2^{r_Z}$), we define a $L$-length binary vector $u\in \mathbb{F}_2^L$, where each element $u_i=1$ means the $i$-th generator is picked up, and ask for the solutions in 
\begin{equation}\label{eq:ux}
    u^T X=0,
\end{equation} up to modulo 2 in addition and multiplication, where $X$ is the ``X" sector of the stabilizer tableau representation.
This restriction means that $u^T Z$ will give a Pauli string only containing $I$ and $Z$ operators, and different $u$ will give different ones.
The number of solutions in Eq.~\eqref{eq:ux} is spanned by the kernel of $X^T$, and can be computed as 
\begin{equation}
    r_Z = \text{dim}\text{Ker}(X^T) = L - \text{rank}(X^T) = L - \text{rank}(X) ,
\end{equation} from the rank-nullity theorem, where $\text{rank}(X)$ is over $\mathbb{F}_2$.
Therefore, the $l_1$-coherence for a stabilizer state can be simply readout from the rank of the $X$ sector over in numerics as
\begin{equation}
    \mathfrak{C}(\rho) = 2^{\text{rank}(X)}-1.
\end{equation} 
Intuitively, the $l_1$-coherence just counts the number of stabilizers that contain $X$ or $Y$ operators in the stabilizer group, since $I$ and $Z$ operators do not contribute to the coherence in the computational basis.

\end{document}